\pdfoutput=1
\documentclass[%
reprint,    
 amsmath,amssymb,
 aps,
 prb,
]{revtex4-2}

\usepackage{graphicx}
\usepackage{dcolumn}
\usepackage{bm}
\usepackage{bbm}
\usepackage{comment}
\usepackage{braket}
\usepackage{siunitx}
\usepackage{subfigure}

\newcommand{\dd}{\mathrm{d}}

\newcommand{\pdv}[2]{\frac{\partial #1}{\partial #2}}

\newcommand{\abs}[1]{\lvert #1\rvert}
\newcommand{\Abs}[1]{\left|#1\right|}

\DeclareMathSymbol{\mhyph}{\mathalpha}{operators}{`-} 

\newcommand{\BZ}{\mathrm{BZ}}	
\newcommand{\UC}{\mathrm{UC}}	
\newcommand{\egap}{\epsilon_{\mathrm{gap}}} 

\usepackage{hyperref}

\begin{document}

\title{
Tunneling spin current in a system with spin degeneracy
}

\author{Yuta Suzuki}
\email{suzuki@vortex.c.u-tokyo.ac.jp}
\affiliation{%
Department of Physics, The University of Tokyo, Bunkyo, Tokyo 113-0033, Japan
}%

\date{\today}

\begin{abstract}
We study theoretically spin current generation from a band insulator with PT symmetry, 
which is associated with Zener tunneling in strong dc electric fields.
Each band in this system is doubly degenerate with opposite spins, 
but spin rotational symmetry is not preserved in general.
We consider the condition for spin current generation 
in connection with the nature of the wave function,
which ultimately depends on a geometric quantity known as the shift vector.
From an analysis of a two-band model, 
we find that the shift vector is necessary for spin current generation
in a PT symmetric system. 
We also present zigzag chain models that have shift vectors, and 
confirm from numerical calculations that a nonzero tunneling spin current occurs 
in spin-degenerate systems.
\end{abstract}

\maketitle


\section{\label{sec:level1}Introduction}

Spin current generation by the application of an electric field has been extensively studied
for crystals with various kinds of symmetries
\cite{Fabian2007,Culcer2007a,Seemann2015}
and for the linear and nonlinear \cite{Hamamoto2017,Kitamura2020b} response regimes.
Even in non-magnetic centrosymmetric crystals, it is possible to generate spin current 
provided spin rotational symmetry is broken by spin--orbit coupling (SOC).
The intrinsic spin Hall effect \cite{Murakami2003} in metals and semiconductors 
is an example of this mechanism.

We can extend this idea to band insulators 
with both inversion and time-reversal symmetries. 
Here, spin current is associated with Zener tunneling \cite{Zener1934} by a strong dc electric field.
This tunneling is considered to be a non-adiabatic process, 
and the Landau--Zener formula \cite{Landau1932,Zener1932,LandauLifshitzQM} is useful for evaluating the
tunneling probability of Bloch electrons across a band gap.
In this formula, the tunneling probability is given by the electric field, band gap, 
and slope of the band dispersion.
The tunneling spin current is then obtained from the difference in tunneling probabilities
between two electrons with opposite spins 
(the definition is given in Eq.~(\ref{eq_spin_current_def}) below).
As mentioned above,
spin current generation by an electric field is also allowed
in the presence of SOC.

\begin{figure}[b]
\includegraphics[bb=0 3 264 168,width=0.9\columnwidth]{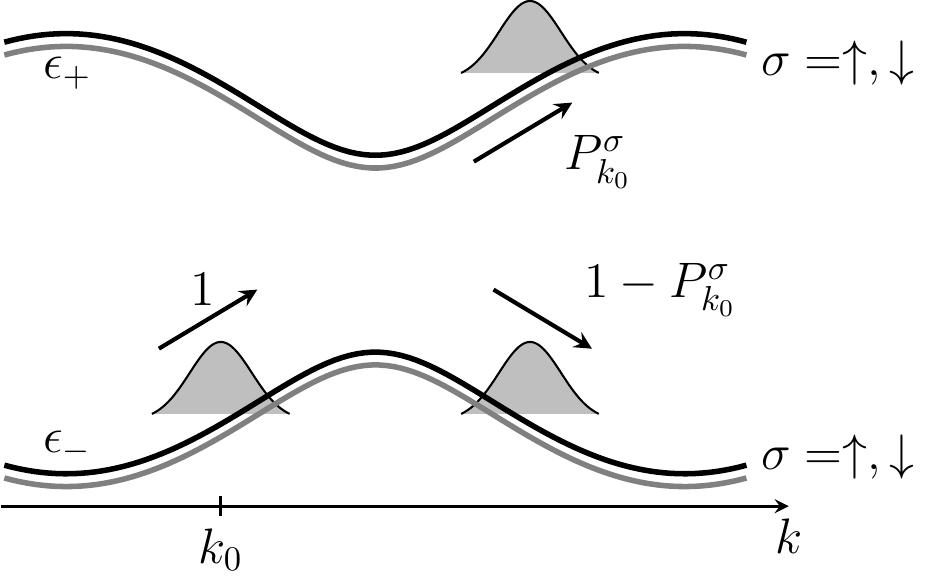}
\caption{Schematic picture of the tunneling process in a system with spin degeneracy.
When an electric field is applied, a wave packet characterized 
by wavenumber $k_0$ and spin $\sigma$
tunnels from the valence band ($\epsilon_-$) to the conduction band ($\epsilon_+$) 
with probability $P^{\sigma}_{k_0}$.
}
\label{fig_tunnel_sketch}
\end{figure}

However, it is not obvious 
how a tunneling spin current arises in crystals with both inversion and time-reversal symmetries.
This can be understood from the band dispersion: 
the combination of the space inversion and the time reversal operations, i.e., a PT operation, transforms 
the wavenumber $k$ and spin $\sigma$ that characterize the energy eigenvalues, as
$(k,\sigma) \to (k, -\sigma)$, and 
each band is doubly degenerate with opposite spins, as shown in Fig.~\ref{fig_tunnel_sketch}.
Thus, focusing on the shape of the dispersions,
it seems that the tunneling probabilities of two electrons with opposite spins are the same,
and that no spin current is allowed in this system.
The same consideration holds for insulators with PT symmetry
that is invariant under the PT operation.

We need to determine the origin of the tunneling spin current in a system with spin degeneracy.
This is a challenge to the conventional idea 
that spin currents hardly arise in spin-degenerate crystals, 
and sheds light on the spin transport in these crystals.

The key to solving the problem 
is the nature of the wave function, rather than degenerate energy bands.
Indeed, even in crystals with both global inversion and time-reversal symmetries,
wave functions with opposite spins can be spatially shifted 
from each other \cite{Zhang2014a,Yuan2019}.
In particular, it has recently been revealed that 
the geometric nature of the wave function is reflected in the tunneling process
\cite{Berry1990,Kitamura2020a,Kitamura2020b,Takayoshi2020} and 
this effect can be reduced to a quantity called the
\emph{shift vector}, which is seen in the asymptotic formula of 
the tunneling probability \cite{Berry1990,Kitamura2020a}.
The shift vector, known for the shift current 
in the photovoltaic effect \cite{Sipe2000,Young2012,Cook2017},
is a gauge-invariant quantity constructed from Berry connections of two bands.
We expect that this quantity plays an important role in the nature of the tunneling phenomenon that
cannot be explained from the band structure.
One example that validates this idea is 
nonreciprocal tunneling in time-reversal systems \cite{Kitamura2020a,Kitamura2020b}.

In this paper, we consider tunneling spin current generation in a PT symmetric system,
in connection with the shift vector.
Our discussion also applies to
crystals with inversion and time-reversal symmetries.

This paper is organized as follows.
In Sec.~\ref{sec_formulation}, we show that
it requires a nonzero shift vector to generate tunneling spin current 
from the analysis of the time-dependent Schr\"{o}dinger equation.
In Sec.~\ref{zigzag_PandT},
we confirm the generation of a tunneling spin current
in a zigzag chain model with a finite shift vector,
based on numerical calculations.
The discussion in this section is generalized to a PT symmetric system in Sec.~\ref{zigzag_PTsym}.
Finally, we provide a summary and conclusions in Sec.~\ref{conclusions}.

\section{\label{sec_formulation}Formulation}
\subsection{\label{subsec:P_j_def}Tunneling probability and current}
We consider a one-dimensional lattice system with PT symmetry.
First, we suppose that no external electric field is applied.
For simplicity, we assume that a certain spin component $\hat{\sigma}$ is 
an exceptionally good quantum number of the system. 
Then, the Hamiltonian in crystal momentum space $\hat{H}(k)$ is described as follows: 
\begin{gather}
 \hat{H}(k) = \hat{H}^{\uparrow}(k)\oplus \hat{H}^{\downarrow}(k),\\
 \hat{H}^{\sigma}(k)\Ket{u^{\sigma}_{n}(k)} = \epsilon^{\sigma}_n(k)\Ket{u^{\sigma}_{n}(k)}
\label{eq_eigenv}
\end{gather}
where the symbols $\uparrow, \downarrow$ correspond, respectively, to the eigenvalues of the spin
$\sigma = +1, -1$.
As we have seen in Sec.~\ref{sec:level1}, PT symmetry leads to 
$\epsilon^{\uparrow}_n(k)= \epsilon^{\downarrow}_n(k)\equiv \epsilon_n(k)$ for any $k$.
From now on, we focus on two gapped bands $n=\pm$ where
$\epsilon_+(k) > \epsilon_-(k)$ holds.

We then consider a dc electric field $E$ applied to this system from time $t=0$.
The Hamiltonian for $t > 0$ is obtained as $\hat{H}^{\sigma}(k-eEt/\hbar)$ 
in accordance with the Peierls substitution.
(In this paper the charge of an electron is written as $-e$.)
Suppose that the valence band $\epsilon^{\sigma}_-(k)$ is occupied 
and the conduction band $\epsilon^{\sigma}_+$ is empty for all $k,\sigma$.
We take a wavenumber $k_0$ from the Brillouin zone (BZ) and define
$\Ket{\Psi^{\sigma}_{k_0}(t)}$ as the time evolution of 
a Bloch state in the valence band $\Ket{u^{\sigma}_{-}(k_0)}$ from $t=0$.
This state satisfies the time-dependent Schr\"{o}dinger equation 
\begin{equation}
 i\hbar \partial_t \Ket{\Psi^{\sigma}_{k_0} (t)} 
= \hat{H}^{\sigma}(k_0(t))\Ket{\Psi^{\sigma}_{k_0} (t)}
\label{TDSE}
\end{equation}
with $k_0(t)= k_0-eEt/\hbar$, and the initial condition 
\begin{equation}
 \Ket{\Psi^{\sigma}_{k_0}(t=0)}=\Ket{u^{\sigma}_{-}(k_0)}~.\label{InitialCond}
\end{equation}
Now, we expand the state with snapshot eigenstates of $\hat{H}^{\sigma}(k_0(t))$ as
\begin{equation}
 \Ket{\Psi^{\sigma}_{k_0} (t)}
 = \sum_{n = \pm } a^{\sigma}_{nk_0}(t) 
e^{i{\gamma}^{\sigma}_{nk_0}(t)}\Ket{u^{\sigma}_{n}(k_0(t))}~.\label{eq_a_expansion_of_Psi}
\end{equation}
Here, we introduced the coefficients $a^{\sigma}_{\pm k_0}$ and 
the sum of the dynamical phase and Berry phase
\begin{equation}
 {\gamma}^{\sigma}_{nk_0}(t)= \int_{k_0}^{k_0(t)}\dd k~\left[
\frac{\epsilon_n(k)}{eE}+A^{\sigma}_{nn}(k)
\right]
\end{equation}
where the Berry connection
$A^{\sigma}_{nm}(k) = \Braket{u^{\sigma}_n(k)|i\partial_k|u^{\sigma}_m(k)}$ appears.
Using these coefficients, the tunneling probability of the Bloch electron is defined as
\begin{equation}
P^{\sigma}_{k_0}(E, t)  = 
\Abs{\Braket{u^{\sigma}_{+}(k_0(t))|\Psi^{\sigma}_{k_0}(t)}}^2
= \Abs{a^{\sigma}_{+k_0}(t)}^2~.
\label{DefTunnelingProb_k0}  
\end{equation}

We can also express
the expectation value of the charge current $j^{\mathrm{c}}= j^{\uparrow} + j^{\downarrow}$ 
and that of the spin current $j^{\mathrm{s}} = \frac{\hbar }{2(-e)}\left(j^{\uparrow} - j^{\downarrow}\right)$
associated with the tunneling.
As in the conventional way, 
$j^{\sigma}$ is given as an expectation value of the velocity operator for all electrons in BZ, and
is expressed as
\begin{equation}
 j^{\sigma}(E,t) = \frac{-e}{L}
\sum_{k_0\in \BZ} 
\Braket{\Psi^{\sigma}_{k_0}(t)|\hat{v}^{\sigma}_{k_0}(t)|\Psi^{\sigma}_{k_0}(t)}~.
\label{eq_current_sum_bz} 
\end{equation}
Here, $L$ is the system size and
$\displaystyle \hat{v}^{\sigma}_{k_0}(t)=\left.\frac{\partial \hat{H}^{\sigma}(k)}{\hbar\partial k}\right|_{k=k_0(t)}$
is the velocity operator.
Note that this definition of the tunneling current is given in \cite{Kitamura2020b},
where the effect of the fermionic heat bath is also considered in the expression.

If we take the limit $L\to \infty$ with periodic boundary conditions,
we find the next expression (shown in Appendix \ref{appendix_expressions_current})
\begin{equation}
j^{\sigma} = \frac{-e}{\hbar}\int_{\BZ}\frac{\dd k_0}{2\pi}~
\left.
\frac{\partial}{\partial k}
\left[
\egap (k) P^{\sigma}_{k_0}(k)
\right]
\right|_{k=k_0(t)}\label{current_a_rep}
\end{equation}
where we introduced $\egap = \epsilon_+ -\epsilon_-$ and considered $P^{\sigma}_{k_0}(t)$ as 
a function of $k=k_0(t)$. 
In particular, the spin current is given as 
\begin{equation}
 j^{\mathrm{s}}=
\int_{\BZ}\frac{\dd k_0}{2\pi}~
\left.
\frac{\partial}{\partial k}
\left[
\egap (k) \frac{P^{\uparrow}_{k_0}(k) -P^{\downarrow}_{k_0}(k)}{2}
\right]
\right|_{k=k_0(t)}
\label{eq_spin_current_def}
\end{equation}
as mentioned in Sec.~\ref{sec:level1}.

\subsection{Explicit expression for the time evolution}
Substituting Eq.~(\ref{eq_a_expansion_of_Psi}) into Eq.~(\ref{TDSE}) gives
\begin{multline}
i\pdv{}{t} a^{\sigma}_{nk_0}(t)
= \frac{eE}{\hbar} \sum_{m (\neq n)} 
\Abs{A^{\sigma}_{nm}(k_0(t))}\\
\cdot e^{i\arg A^{\sigma}_{nm}(k_0)}e^{-i\Delta^{\sigma}_{nm}(k_0(t),k_0)}
a^{\sigma}_{mk_0}(t)
\label{TimeEvolution_a(t)}
\end{multline}
where we introduced
\begin{equation}
 \Delta^{\sigma}_{nm}(k,k_0)= 
\int_{k_0}^{k}\dd k'~
\frac{\epsilon_n(k')-\epsilon_m(k') + eE\cdot R^{\sigma}_{nm}(k')}{eE}~.\label{eq_def_Delta} 
\end{equation}
Here, 
\begin{equation}
R^{\sigma}_{nm}(k)= A^{\sigma}_{nn}(k) -A^{\sigma}_{mm}(k) -\partial_k\arg A^{\sigma}_{nm}(k) 
\end{equation}
is called the \emph{shift vector}.

As Eq.~(\ref{InitialCond}) gives
$\left.(a^{\sigma}_{+k_0}, a^{\sigma}_{-k_0})\right|_{t=0} = (0,1)$, 
we get a solution of Eq.~(\ref{TimeEvolution_a(t)}) formally expressed as 
\begin{widetext}
\begin{equation}
 \begin{pmatrix}
 a^{\sigma}_{+k_0}(t)e^{-i\arg A^{\sigma}_{+-}(k_0)}\\
 a^{\sigma}_{-k_0}(t)
\end{pmatrix}
= \hat{T}
\exp\left[
i\int_{0}^{t}
\left.
\abs{A^{\sigma}_{+-}(k)}
\begin{pmatrix}
 0 & e^{-i{\Delta^{\sigma}_{+-}(k,k_0)}}\\
e^{+i{\Delta^{\sigma}_{+-}(k,k_0)}} & 0
\end{pmatrix}
\right|_{k=k_0(t')}
\frac{-eE}{\hbar}\dd t'
\right]
\begin{pmatrix}
 0\\  1
\end{pmatrix}
~.\label{eq_Tprod}
\end{equation} 
\end{widetext}
Here, $\hat{T}$ is the time ordering operator.
This formula shows that the information contained in the Hamiltonian 
is based on three quantities $\egap (k)$, $\Abs{A^{\sigma}_{+-}(k)}$ and $R^{\sigma}_{+-}(k)$ 
that are all found to be gauge-invariant. 
Note that $R^{\sigma}_{+-}(k)$ is interpreted as being 
the difference in the intracell coordinate of the Bloch electron between 
the valence and conduction bands \cite{Morimoto2016}.
From this perspective, 
the spatial shift $R^{\sigma}_{+-}$ yields the difference of the electrostatic potential between the two bands
$eE\cdot R^{\sigma}_{+-}$, 
which changes the energy to cross the band gap from $\egap$ to $\egap + eE\cdot R^{\sigma}_{+-}$.
We can find this term in Eq.~(\ref{eq_def_Delta}).
In real space, this effect
appears as a correction to the depth of the tunnel barrier \cite{Kitamura2020a}.

\subsection{\label{subsec:Orign_spin_dep}Origin of the spin dependence}
The Berry connection and shift vector satisfy
\begin{equation}
\Abs{A^{\uparrow}_{+-}(k)}=\Abs{A^{\downarrow}_{+-}(k)}, \quad
R^{\uparrow}_{+-}(k)= -R^{\downarrow}_{+-}(k)
\label{ARRelationwPTsym}
\end{equation}
in the PT symmetric system.
In a system with both inversion and time-reversal symmetries,
as a special case of PT symmetry, 
\begin{equation}
\Abs{A^{\sigma}_{+-}(-k)}=\Abs{A^{\sigma}_{+-}(+k)},\quad
R^{\sigma}_{+-}(-k)= -R^{\sigma}_{+-}(+k)
\label{ARRelationwPandTsym}
\end{equation}
also holds for $\sigma = \uparrow, \downarrow$
in addition to Eq.~(\ref{ARRelationwPTsym}).
These derivations are given in Appendix \ref{appendix_symmetry}.
Note that the shift vector remains finite in general even in 
such a highly symmetric system, though
it vanishes if the system has spin rotational symmetry, in addition \cite{Kitamura2020a}.

We now consider spin dependence of the tunneling probability
in a PT symmetric system.
Equation~(\ref{DefTunnelingProb_k0}) shows that 
the spin dependence results from 
the difference between $\Abs{a^{\uparrow}_{+k_0}}^2$ and $\Abs{a^{\downarrow}_{+k_0}}^2$.
Here, the formula for $a^{\sigma}_{nk_0}$ is given in Eq.~(\ref{eq_Tprod}) and
its spin dependence is reduced to that of $\Abs{A^{\sigma}_{+-}(k)}$ and $R^{\sigma}_{+-}(k)$ 
in $\Delta^{\sigma}_{+-}(k, k_0)$.
In addition, $\Abs{A^{\sigma}_{+-}(k)}$ is spin-independent, as shown in Eqs.~(\ref{ARRelationwPTsym}).
Thus, the spin dependence of the tunneling probability is dependent only on the shift vector that changes its sign depending on its spin.
We can also explain this result in terms of the energy to cross the band gap $\egap + eE\cdot R^{\sigma}_{+-}$.
This energy is different between electrons with opposite spins,
which leads to the difference in tunneling probability.

Furthermore, we can consider the condition of the spin current generation.
If $R^{\sigma}_{+-}= 0$ holds, we then find that 
$P^{\uparrow}_{k_0}(t) = P^{\downarrow}_{k_0}(t)$ and 
$j^{\mathrm{s}} = 0$ hold because of Eq.~(\ref{eq_spin_current_def}).
This result implies that even in a system with spin degeneracy,
tunneling spin current can be generated unless $R^{\sigma}_{+-}= 0$.

\section{Zigzag chain model with both inversion and time-reversal symmetries}
\label{zigzag_PandT}

In this section, 
we construct a simulation model with both inversion and time-reversal symmetries
and derive a two-band Hamiltonian to show that tunneling spin current is induced by an electric field.
Note that the model has previously been introduced in \cite[Sec.~7.2]{Kusunose2019}, except that 
a parameter $v$ is added in the current study, as will be detailed later.
In our model, we show that the shift vector remains finite since $v\neq 0$, and
we confirm that the shift vector plays an important role in the spin current generation.

\subsection{Tight-binding Hamiltonian}
\begin{figure}[b]
\subfigure[]{%
\includegraphics[bb=3 0 325 121,width=0.9\columnwidth]{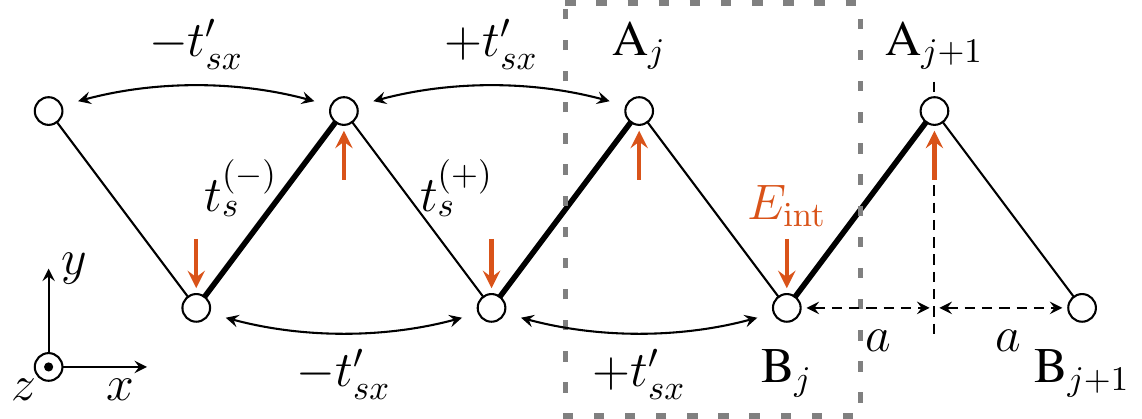}
\label{fig_zigzagPandT}
}%
\\
\subfigure[]{%
\includegraphics[bb=0 0 217 125,width=0.9\columnwidth]{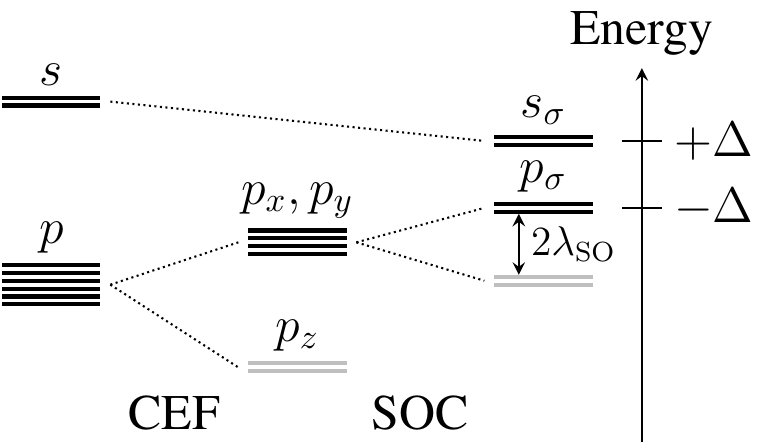}
\label{fig_ene_spl}
}%
\caption{(Color online) 
Zigzag chain model and energy splitting diagram at each site.
(a) The dashed gray rectangle shows 
a unit cell ($j$-th), with two sublattices labeled A and B and lattice spacing $a$.
The red vertical arrows denote a staggered internal electric field and the 
thick/thin solid lines denote nearest-neighbor hoppings (between two $s$ orbitals),
while the arcs denote next nearest-neighbor hoppings (between $s$ and $p_x$ orbitals).
(b) The degeneracy here includes the spin degree of freedom.
CEF and SOC stand for crystalline electric field and spin--orbit coupling, respectively.
}
\label{fig_zigzag_ene_spl}
\end{figure}

Let us consider a one-dimensional zigzag chain, shown
in Fig.~\ref{fig_zigzagPandT}, with $s, p$ orbitals at each atomic site.
This chain satisfies the following three conditions: 
(i) it has no magnetic orders and has time-reversal symmetry; 
(ii) it has inversion symmetry and the inversion center is located at the middle of the neighboring A and B sites;
(iii) it has reflection symmetry in a plane perpendicular to the $z$ axis,
while it breaks reflection symmetry in a plane perpendicular to the $x$ axis.
We also introduce a crystalline electric field (CEF) that is consistent with 
the conditions above.

Taking all this into account, we focus on 
the energy levels at each site as shown in Fig.~\ref{fig_ene_spl}.
First, the CEF lifts the six-fold degenerate $p$ orbitals into the energy levels $(p_x, p_y)$ and
$p_z$. 
Then, we consider the splitting of the $p_x, p_y$ levels due to the SOC, which is given as
$\hat{{H}}_{\mathrm{SO}}=\lambda_{\mathrm{SO}} \hat{\bm{l}}\cdot \hat{\bm{\sigma}}$.
Here, $\hat{\bm{l}}$ and $\hat{\bm{\sigma}}$ are the orbital and spin angular momentum, respectively.
We find that $\hat{{l}}_x, \hat{{l}}_y$ have no effect on the splitting
because the $p_x, p_y$ orbitals are related to the eigenstates of $\hat{{l}}_z$ through
\begin{equation}
\ket{\pm 1} \equiv \ket{l = 1, m = \pm 1}
= \frac{1}{\sqrt{2}}\left(\ket{p_x}\pm i\ket{p_y}\right)
\end{equation}
and $\Braket{\pm 1|\hat{{l}}_{x, y}|\pm 1}=0$ holds.
Therefore, 
the term $\lambda_{\mathrm{SO}}\hat{{l}}_z\cdot \hat{{\sigma}}_z$ lifts
the degeneracy into two Kramers pairs,
$(\ket{+1, \uparrow}, \ket{-1, \downarrow})$ and 
$(\ket{+1, \downarrow}, \ket{-1, \uparrow})$.
We assume that the former pair
\begin{equation}
 \ket{p_{\sigma}} \equiv \frac{1}{\sqrt{2}}\left(\ket{p_x,\sigma}+ i\sigma\ket{p_y,\sigma}\right)
\quad (\sigma = \sigma_z = \pm 1)\label{eq_Def_psigma}
\end{equation}
is at an energy level close to that of the $s$ orbitals $\ket{s_{\sigma}}\equiv \ket{s,\sigma}$, 
as shown in Fig.~\ref{fig_ene_spl}.
For simplicity, we take the energy levels of $\ket{s_{\sigma}}, \ket{p_{\sigma}}$
to be $+\Delta, -\Delta$.

We then focus on the onsite parity mixing due to the internal electric field as shown in Fig.~\ref{fig_zigzagPandT}, 
which appears when the inversion center is not located at a site.
To see this, we introduce the local electrostatic potential $\hat{V}_{\mathrm{in}}$ as follows:
\begin{equation}
 \hat{V}_{\mathrm{in}}({y}) =
\begin{cases}
 -E_{\mathrm{in}}{y} & \text{near A sites}\\
 +E_{\mathrm{in}}{y} & \text{near B sites}
\end{cases} 
\end{equation}
where the inversion center is placed at $y=0$. 
Here, $\hat{V}_{\mathrm{in}}$ satisfies the global inversion symmetry 
and meets the conditions (i)--(iii). 
Then, the mixing $\Braket{s|(-e)\hat{V}_{\mathrm{in}}|p_y}$ occurs 
at each site
and we find from Eq.~(\ref{eq_Def_psigma}) that
\begin{equation}
 \braket{s_{\sigma A}|(-e)\hat{V}_{\mathrm{in}}|p_{\sigma A}}
= -\braket{s_{\sigma B}|(-e)\hat{V}_{\mathrm{in}}|p_{\sigma B}} = i\sigma \phi.
\end{equation}
Here, we introduced a constant $\phi$.

We also introduce the transfer integrals $t^{(\pm)}_{s}$ and $t'_{sx}$ as shown in Fig.~\ref{fig_zigzagPandT}
The signs with $t'_{sx}$ are determined by the direction dependence of $\ket{s}, \ket{p_x}$.
It is also natural to assume that $t^{(+)}_s$ and $t^{(-)}_s$ have the same sign in this model.

Thus, we get the tight-binding Hamiltonian
\begin{align}
&\hat{\mathcal{H}}^{\mathrm{s\mhyph p}}
=\sum_{j,\sigma,\tau}
\Delta \left(
\hat{s}^{\dagger}_{j\tau\sigma}~\hat{s}_{j\tau\sigma}
-\hat{p}^{\dagger}_{j\tau\sigma}~\hat{p}_{j\tau\sigma}
\right)\nonumber\\
&+\sum_{j, \sigma}(i\sigma\phi)\left(
\hat{s}^{\dagger}_{j\mathrm{A}\sigma}~\hat{p}_{j\mathrm{A}\sigma}
-\hat{s}^{\dagger}_{j\mathrm{B}\sigma}~\hat{p}_{j\mathrm{B}\sigma}
\right)\nonumber\\
&+\sum_{j,\sigma}
\left(
t^{(+)}_s\hat{s}^{\dagger}_{j\mathrm{A}\sigma}~\hat{s}_{j,\mathrm{B}\sigma}
+t^{(-)}_s\hat{s}^{\dagger}_{j\mathrm{A}\sigma}~\hat{s}_{j-1,\mathrm{B}\sigma}
\right)\nonumber\\
&+\sum_{j,\sigma,\tau}
t'_{sx}\left(
\hat{s}^{\dagger}_{j\tau\sigma}~\hat{p}_{j+1,\tau\sigma}
-\hat{s}^{\dagger}_{j\tau\sigma}~\hat{p}_{j-1,\tau\sigma}\right)+ \mathrm{H.c.} 
\end{align}
where
$\hat{s}^{\dagger}_{j\tau\sigma}, \hat{s}_{j\tau\sigma}$,
 $\hat{p}^{\dagger}_{j\tau\sigma}, \hat{p}_{j\tau\sigma}$
are electron creation and annihilation operators 
on the $j$-th site, labeled with orbital ($s, p$), spin ($\sigma = \uparrow, \downarrow$)
and sublattice ($\tau = \mathrm{A},\mathrm{B}$).

In the crystal momentum space,
the Hamiltonian is expressed as follows:
\begin{align}
 \hat{\mathcal{H}}^{\mathrm{s\mhyph p}} &= \sum_{k\sigma} 
\hat{C}^{\dagger}_{k\sigma}{H}^{\mathrm{s\mhyph p}}_{k\sigma}\hat{C}_{k\sigma},\label{eq_Sigma_H(k)}\\
{H}^{\mathrm{s\mhyph p}}_{k\sigma}
&=
\begin{pmatrix}
 \Delta &{\alpha}_k  & {\beta}^{(+)}_{k\sigma} & 0\\
 {{\alpha}_k}^{\ast} &\Delta &0 & {\beta}^{(-)}_{k\sigma}\\
 {{\beta}^{(+)}_{k\sigma}}^{\ast} &{0} &-\Delta & 0 \\
 {0} &{{\beta}^{(-)}_{k\sigma}}^{\ast}&{0} & -\Delta
\end{pmatrix}
~.
\end{align}
Here, we introduced
\begin{equation}
 {\alpha}_k = t^{(+)}_se^{+ika}+t^{(-)}_se^{-ika},\quad
{\beta}^{(\pm)}_{k\sigma} = 2it'_{sx}\sin(2ka)\pm i\phi\sigma
\end{equation}
and $\hat{C}^{\dagger}_{k\sigma} = (\hat{s}^{\dagger}_{kA\sigma},\ 
\hat{s}^{\dagger}_{kB\sigma},\ \hat{p}^{\dagger}_{kA\sigma},\ \hat{p}^{\dagger}_{kB\sigma})$
obtained from the Fourier transformation of $\hat{s}^{\dagger}_{j\tau\sigma}, \hat{p}^{\dagger}_{j\tau\sigma}$.
Let us now assume that $\phi, t^{(\pm)}_s, t'_{sx}\ll \Delta$
and derive the effective Hamiltonian for $\ket{s_{\sigma}}$.
We divide 
$4\times 4$ matrix ${H}^{\mathrm{s\mhyph p}}_{k\sigma}$ 
into $2\times 2$ blocks as
\begin{equation}
{H}^{\mathrm{s\mhyph p}}_{k\sigma}= 
\begin{pmatrix}
 {H}^s_{k\sigma} & {H}^{1}_{k\sigma}\\
 {{H}^{1}_{k\sigma}}^{\dagger}&{H}^p_{k\sigma}
\end{pmatrix} 
\end{equation}
and obtain the effective Hamiltonian
\begin{equation}
 \tilde{{H}}^s_{k\sigma}
\simeq {H}^s_{k\sigma} + \frac{1}{2\Delta }{H^1_{k\sigma}}{H^1_{k\sigma}}^{\dagger}
= H^{\sigma}(k)+ (\text{scalar})
\end{equation}
where we can neglect the scalar term since it has no effect on the tunneling process.
We find that
\begin{align}
&H^{\sigma}(k)=
\begin{pmatrix}
 -u\sigma \sin (ka) \cos (ka) & {w}\cos (ka) -iv\sin (ka)\\
 {w}\cos (ka)+iv\sin (ka)& u\sigma \sin (ka) \cos (ka)
\end{pmatrix}
\nonumber\\
&= \bm{d}^{\sigma}(k)\cdot\bm{\tau} \label{eq_Hamiltonian_PandT}
\end{align}
with Pauli matrices $\bm{\tau}= \left(\tau_x, \tau_y, \tau_z\right)$.
This two-band Hamiltonian is characterized by the following:
\begin{gather}
 u = \frac{-4t'_{sx}\phi}{\Delta},\quad 
 v = t^{(-)}_{s} - t^{(+)}_{s}, \quad
 {w}= t^{(+)}_{s} + t^{(-)}_{s},\label{eq_uvt_expression}\\
\bm{d}^{\sigma}(k) =({w}\cos (ka), v\sin (ka), -\sigma u\sin (ka)\cos (ka))~.\label{eq_component_hamiltonian}
\end{gather}

Let us examine each component of $\bm{d}^{\sigma}(k)$.
The first component always remains finite since $t^{(+)}_s$ and $t^{(-)}_s$ have the same sign, as mentioned above.
The second component with $v$, however, disappears if the model recovers
reflection symmetry in a plane perpendicular to the $x$ axis, since $t^{(+)}_s = t^{(-)}_s$ holds in this case.
Thus, we can consider this component to be the result of reflection symmetry breaking.
We also find that 
the third component of $\bm{d}^{\sigma}(k)$ emerges due to the lack of local inversion symmetry,
since $\phi= 0$ holds if an inversion center can be placed at each lattice site.
This term is known as the antisymmetric spin--orbit interaction
in lattice structures such as zigzag \cite{Yanase2014,Sugita2015}, honeycomb and diamond structure, which are all
globally centrosymmetric systems but break the inversion symmetry at lattice sites \cite{Hayami2016}.
Note that it is easy to see that this component corresponds to SOC
when the SOC splitting is small enough to satisfy $\lambda_{\mathrm{SO}} \ll \Delta$. 
In this case, 
we should take into account the additional energy level just below $\ket{p_{\sigma}}$ 
shown in Fig.~\ref{fig_ene_spl}. The parameter $u$ is then modulated as 
\begin{equation}
 u=\frac{-4t'_{sx}\phi}{\Delta}
\to -4t'_{sx}\phi
\left(
\frac{1}{\Delta} -\frac{1}{\Delta + \lambda_{\mathrm{SO}}}
\right)
\simeq 
\frac{-4t'_{sx}\phi\lambda_{\mathrm{SO}}}{\Delta^2}
\end{equation}
and is indeed proportional to $\lambda_{\mathrm{SO}}$.

Now, we check if the shift vector in this model remains finite.
From the Hamiltonian (\ref{eq_Hamiltonian_PandT}), we can express it as follows:
\begin{multline}
R^{\sigma}_{+-}(k) = \frac{
(\bm{d}^{\sigma}\times \partial_k\bm{d}^{\sigma})\cdot (\partial_k^2\bm{d}^{\sigma})
}{
(\bm{d}^{\sigma}\times \partial_k\bm{d}^{\sigma})^2
}\sqrt{{(\bm{d}^{\sigma})}^2}\\
= \frac{3}{2}uv{w}\sin (2k)
\frac{
\sqrt{u^2\sin^2k\cos^2k+v^2\sin^2k+{w}^2\cos^2k}
}{v^2{w}^2 + u^2{w}^2\cos^6k + u^2v^2\sin^6k}\sigma
\label{eq_Rrep} 
\end{multline}
where we used the formula given in \cite{Kitamura2020a} and take $a=1$. 
Equation~(\ref{eq_Rrep}) shows that $R^{\sigma}_{+-}(k)\neq 0$ holds when $u, v, {w}\neq 0$.

\subsection{Numerical calculations}
\begin{figure*}
\subfigure[]{%
\includegraphics[bb=2 3 508 388,width=0.7\columnwidth]{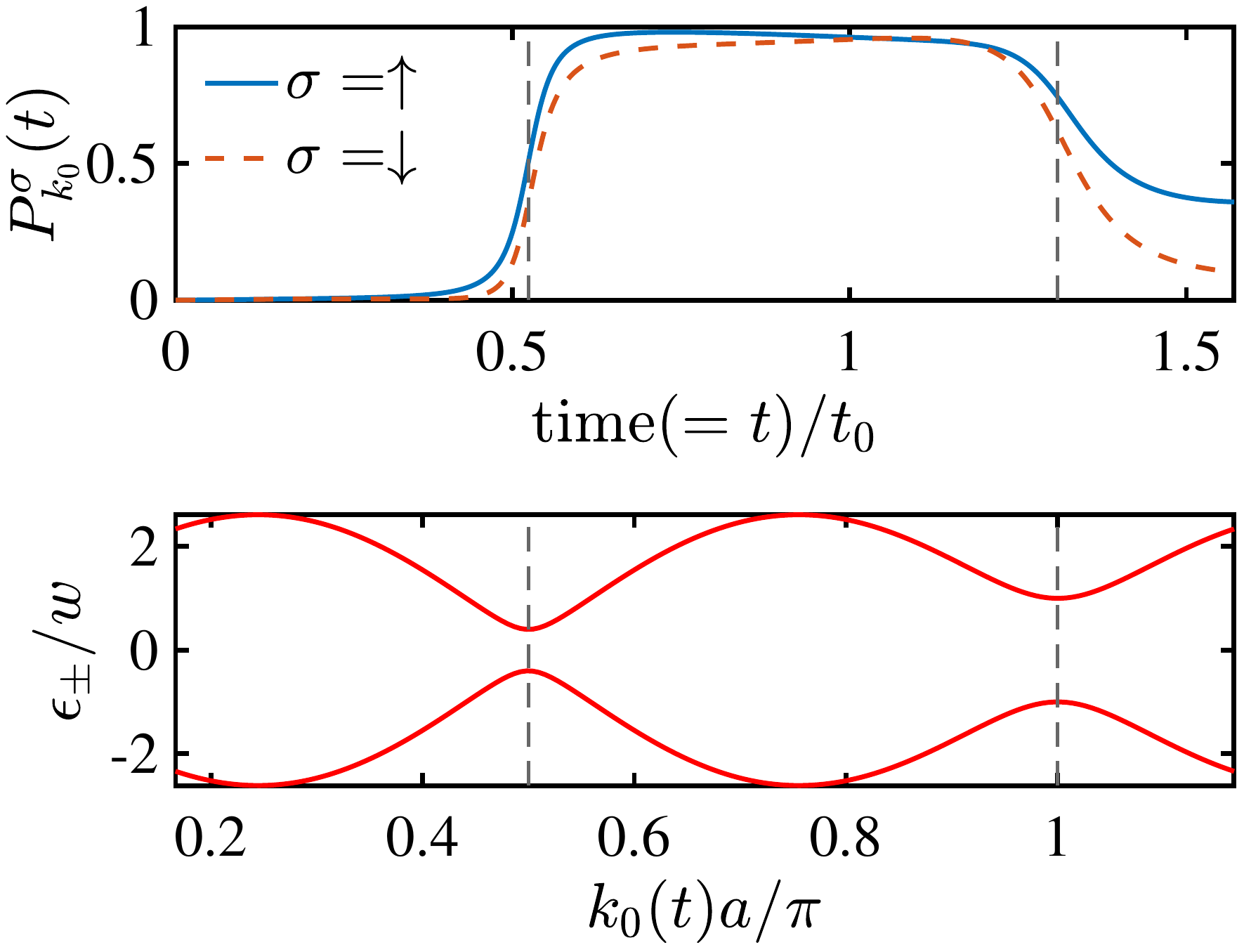}
\label{fig_tunnel_prob_2rows}}%
\quad
\subfigure[]{%
\includegraphics[bb=2 2 519 398,width=0.7\columnwidth]{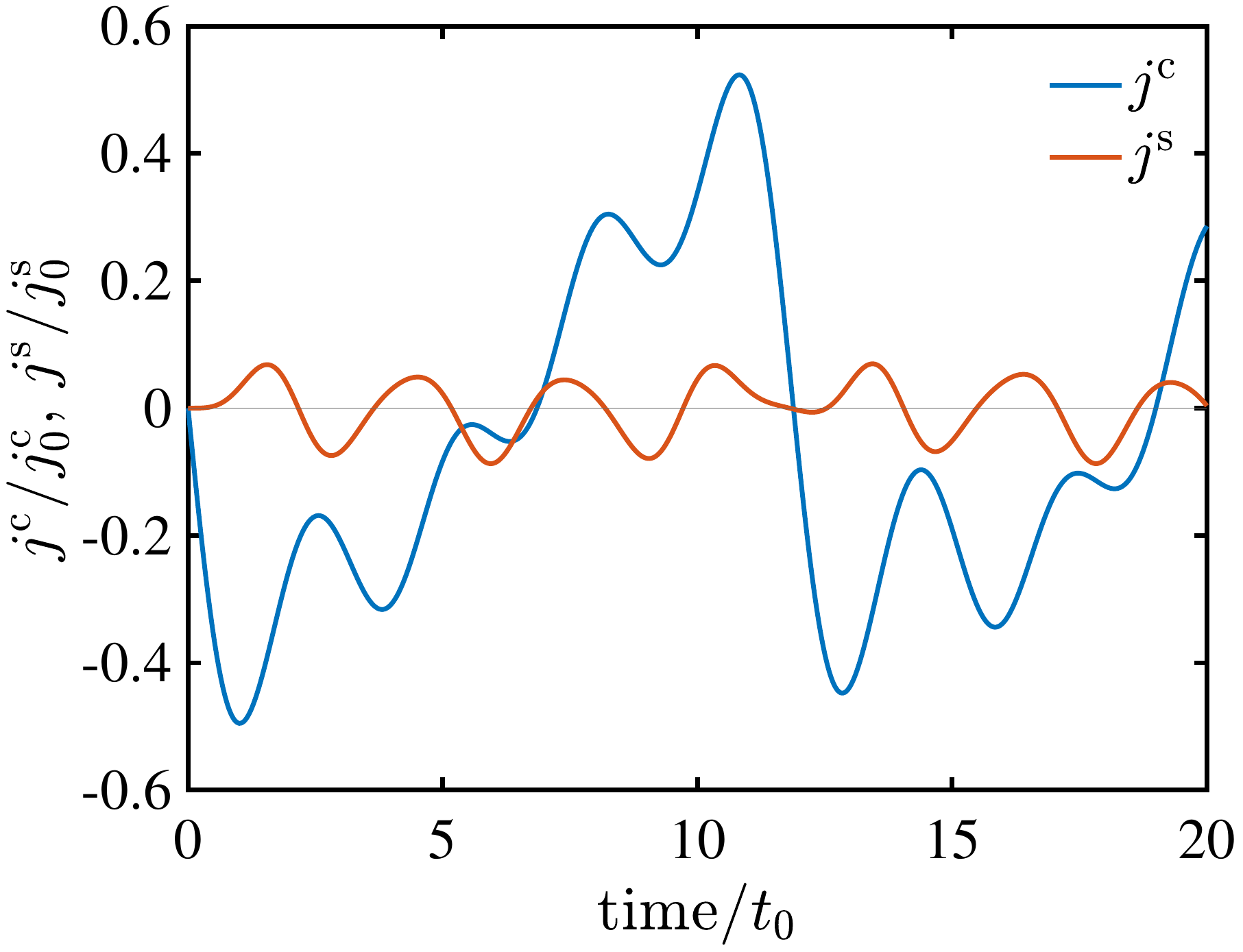}
\label{fig_charge_spin_current}}%
\\
\subfigure[]{%
\includegraphics[bb=3 2 500 415,width=0.7\columnwidth]{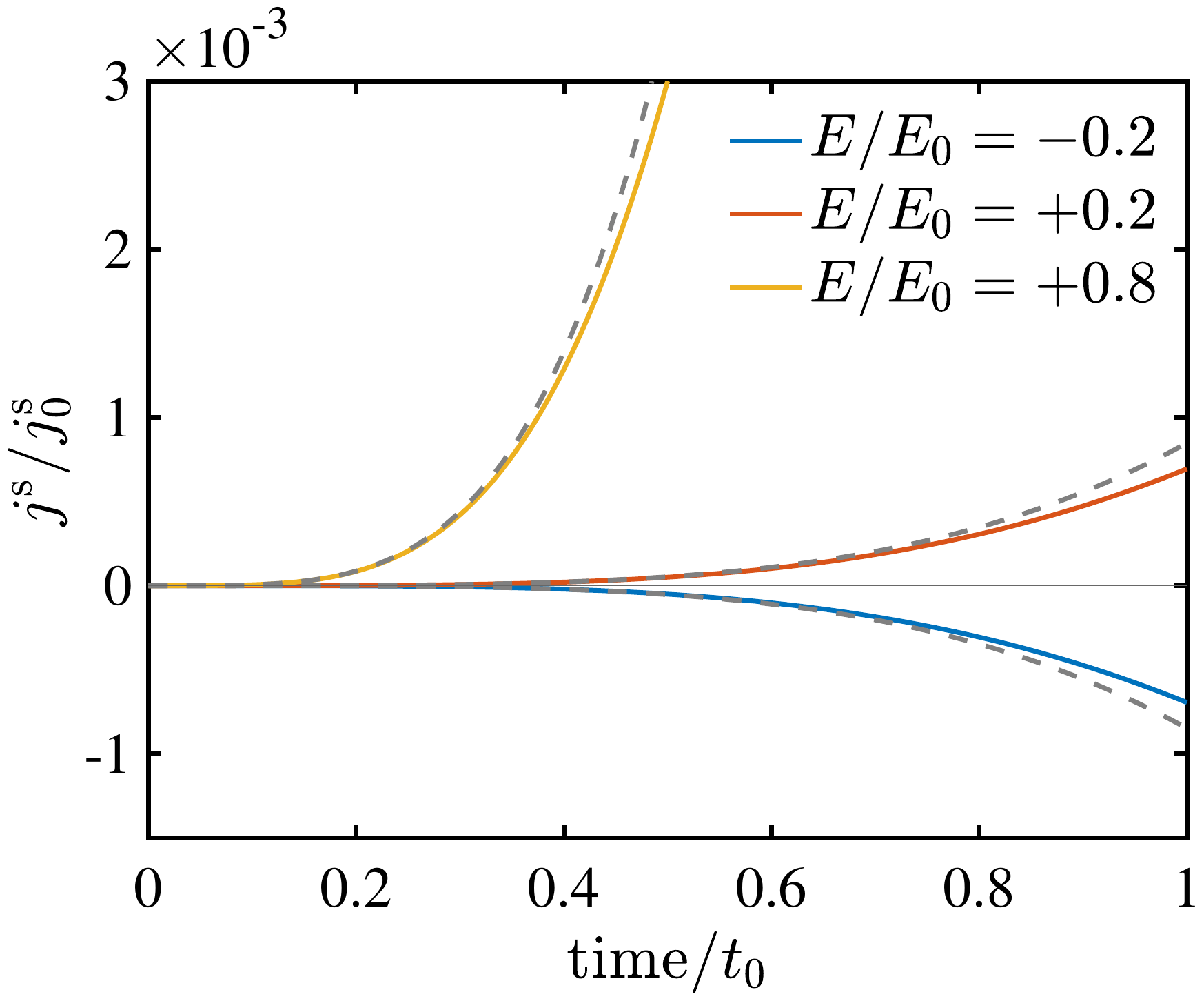}
\label{fig_js_dE}}%
\quad
\subfigure[]{%
\includegraphics[bb=3 2 510 415,width=0.7\columnwidth]{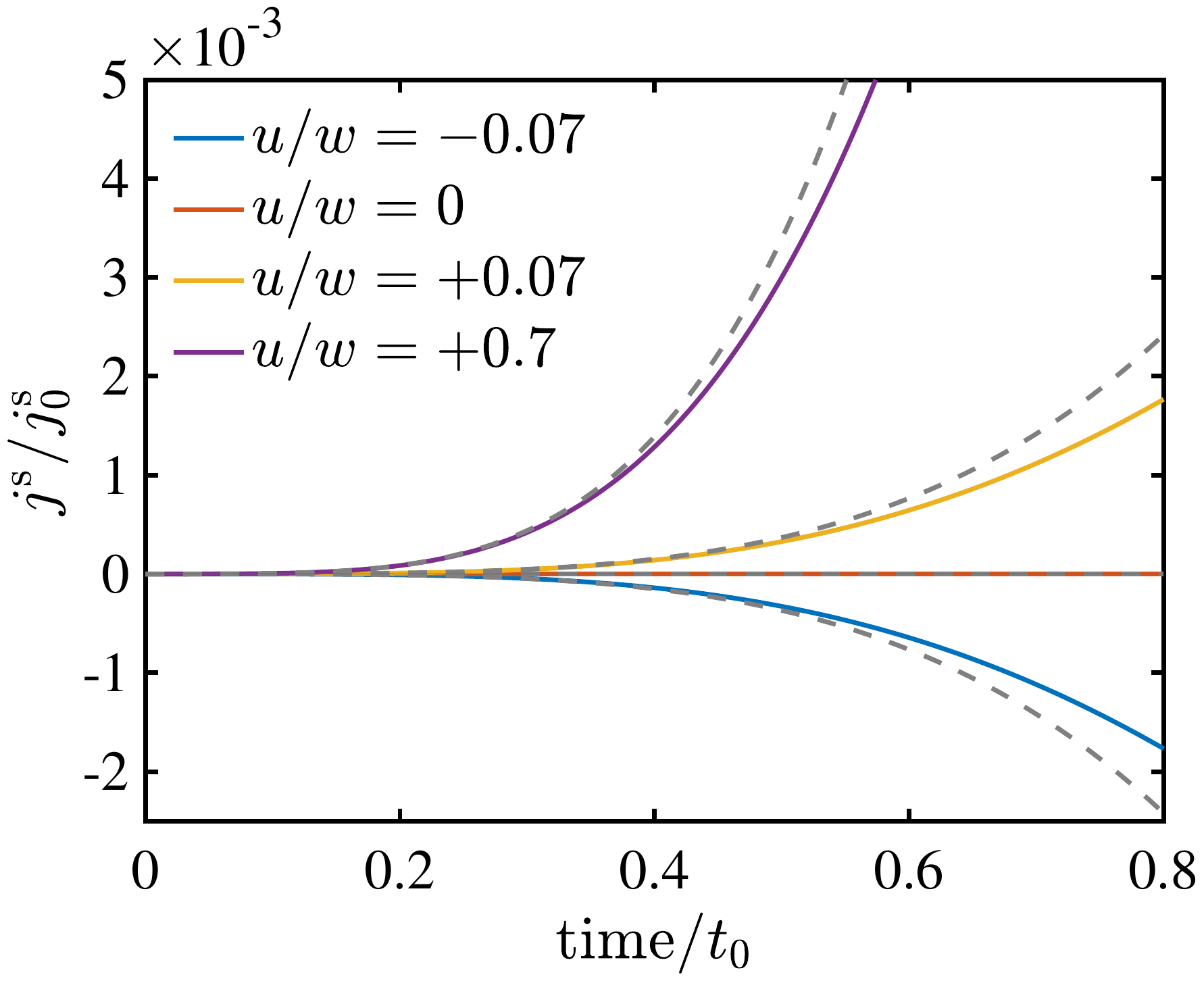}
\label{fig_js_du}}%
\caption{(Color online) 
Numerical results for time variation of tunneling probability and charge/spin current in the 
zigzag chain model with both inversion and time-reversal symmetries. 
(a) Tunneling probability (upper row) across the energy levels (lower row) with the parameters $(k_0a, u/{w}, v/{w}, E/E_0) = (\pi/6, 5, 0.4, -2)$.
The dashed vertical lines show where the band gap is locally minimum.
(b) Charge and spin currents with the parameters $(u/{w}, v/{w}, E/E_0)=(0.7, 0.4, 0.8)$.
(c) Electric field dependence of the spin current shortly after application of electric field with the parameters $(u/{w}, v/{w})=(0.7, 0.4)$.
The dashed curves represent $j^{\mathrm{s}}(E, t)$ from the analytical formula (\ref{eq_approx_js_PandT}).
(d) Spin current shortly after the application of an electric field for different values of
antisymmetric SOC parameter $u$, with the parameters $( v/{w}, E/E_0)=(0.4, 0.8)$.
The dashed curves represent $j^{\mathrm{s}}(E, t)$ from the analytical formula (\ref{eq_approx_js_PandT}).
}
\label{4figs_PandT}
\end{figure*}

From the Hamiltonian (\ref{eq_Hamiltonian_PandT}), 
the behaviors of $P^{\sigma}_{k_0}(t)$, $j^{\mathrm{c}}(E, t)$ and $j^{\mathrm{s}}(E, t)$ 
are obtained numerically as shown in Fig.~\ref{4figs_PandT}.
Here, 
we introduced $t_0=\hbar/{w}$, $E_0 = {w}/(ea)$, $j^{\mathrm{c}}_0 = (-e){w}/\hbar$ and $j^{\mathrm{s}}_0 = {w}/2$
as a typical time length, electric field, charge current and spin current, respectively.

In Fig.~\ref{fig_tunnel_prob_2rows}, 
we can see that the probabilities with opposite spins are different from each other,
which is expected 
since $u\neq 0$ breaks spin rotational symmetry in our model.
Note that $P^{\sigma}_{k_0}(t)$ has a steep rise and fall 
in the vicinity of $k=k_0(t)$ where the band gap is locally minimum. 
This behavior is characteristic of tunneling.
We also find from Fig.~\ref{fig_charge_spin_current} that 
a nonzero spin current is associated with an electric field in this model.
This spin current generation purely originates from a nonzero shift vector, i.e., 
the geometric effect of the wave function, since each band is doubly degenerate with opposite spins.

Figure~\ref{fig_charge_spin_current} also shows an oscillating behavior in the charge and spin currents, 
the reason for which can be explained as follows: 
the Hamiltonian of this model 
$\hat{H}^{\sigma}(k_0(t))$ is time-periodic with period $T\equiv 2\pi\hbar/(e\Abs{E}a)$, and
we can consider that the Schr\"{o}dinger equation (\ref{TDSE}) describes a kind of Rabi cycle driven by
an external field with period $T$. Thus, we see
an oscillation in the population $\Abs{a^{\sigma}_{nk_0}(t)}^2$ 
and that of the currents $j^{\mathrm{s}}(t)$.
For more details, see Appendix \ref{appendix_oscillation}.
Here, $T$ corresponds to the period of the Bloch oscillation, though
it is hard to observe the oscillation in experiments since $T$ is large.
Thus, we focus on the time period shortly after application of the electric field, i.e., $t\ll t_0$.

In this region, the tunneling spin current can be approximated as
\begin{multline}
j^{\mathrm{s}}(E, t)/j^{\mathrm{s}}_0 \\
\simeq \frac{-5}{12\pi} 
\left(\frac{1}{a^3{{w}}^2}\int_{\BZ}\dd k~ 
\egap(\partial_k \egap){\Abs{A^{\uparrow}_{+-}}}^2 R^{\uparrow}_{+-}
\right)\\
\cdot \left(\frac{E}{E_0}\right)^{3}\left(\frac{t}{t_0}\right)^4~,\label{eq_approx_js_PandT}  
\end{multline}
and is scaled by the integral containing the shift vector $R^{\uparrow}_{+-}$.
This expression is derived in Appendix \ref{sec_approx_tunneling_current}.
In Fig.~\ref{fig_js_dE}, this formula fits the numerically obtained $j^{\mathrm{s}}(t)$ for each value of $E/E_0$.
The relation 
$j^{\mathrm{s}}(-E)=-j^{\mathrm{s}}(+E)$ also holds in both Fig.~\ref{fig_js_dE} and Eq.~(\ref{eq_approx_js_PandT}), 
which is expected from the inversion symmetry of the model.
 
As we mentioned in Sec.~\ref{subsec:Orign_spin_dep}, this spin current disappears when the shift vector goes to zero.
This means that $j^{\mathrm{s}} = 0$ holds if the antisymmetric SOC vanishes ($u=0$) or 
the reflection symmetry is restored ($v=0$) in our zigzag chain model.
The curve of $u = 0$ in Fig.~\ref{fig_js_du} demonstrates this behavior.
Moreover, Fig.~\ref{fig_js_du} shows a trend of the spin current increasing as $u$ moves away from zero.
In the Supplemental Material, we check that the same trend holds for the parameter $v$.
These results and the integral in Eq.~(\ref{eq_approx_js_PandT}) indicate that 
as long as $u, v$ are small enough, 
the shift vector is increased by the antisymmetric SOC and reflection symmetry breaking, 
which enhances the tunneling spin current in the zigzag chain model. 

\section{General zigzag chain model with PT symmetry}
\label{zigzag_PTsym}

\begin{figure*}
\subfigure[]{%
\includegraphics[bb=3 0 325 121,width=0.9\columnwidth]{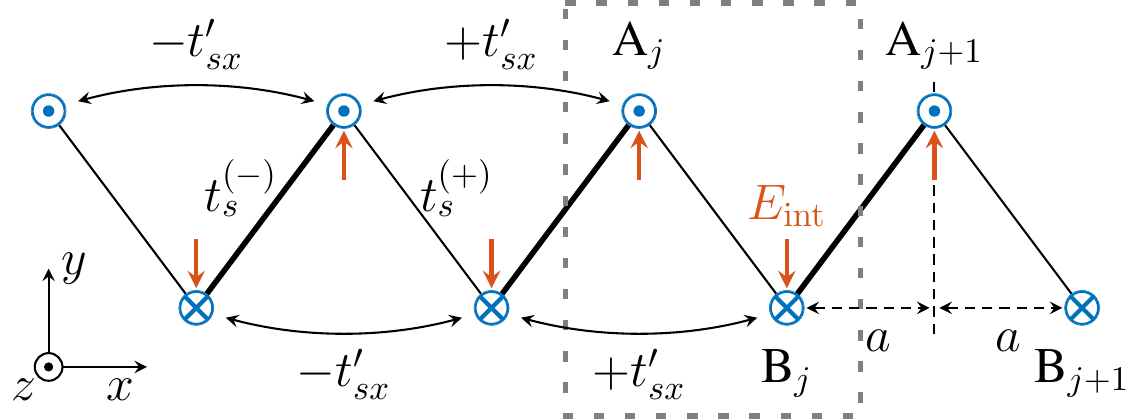}
\label{fig_zigzag_chain_PTsym}}%
\\
\subfigure[]{%
\includegraphics[bb=2 2 519 398,width=0.7\columnwidth]{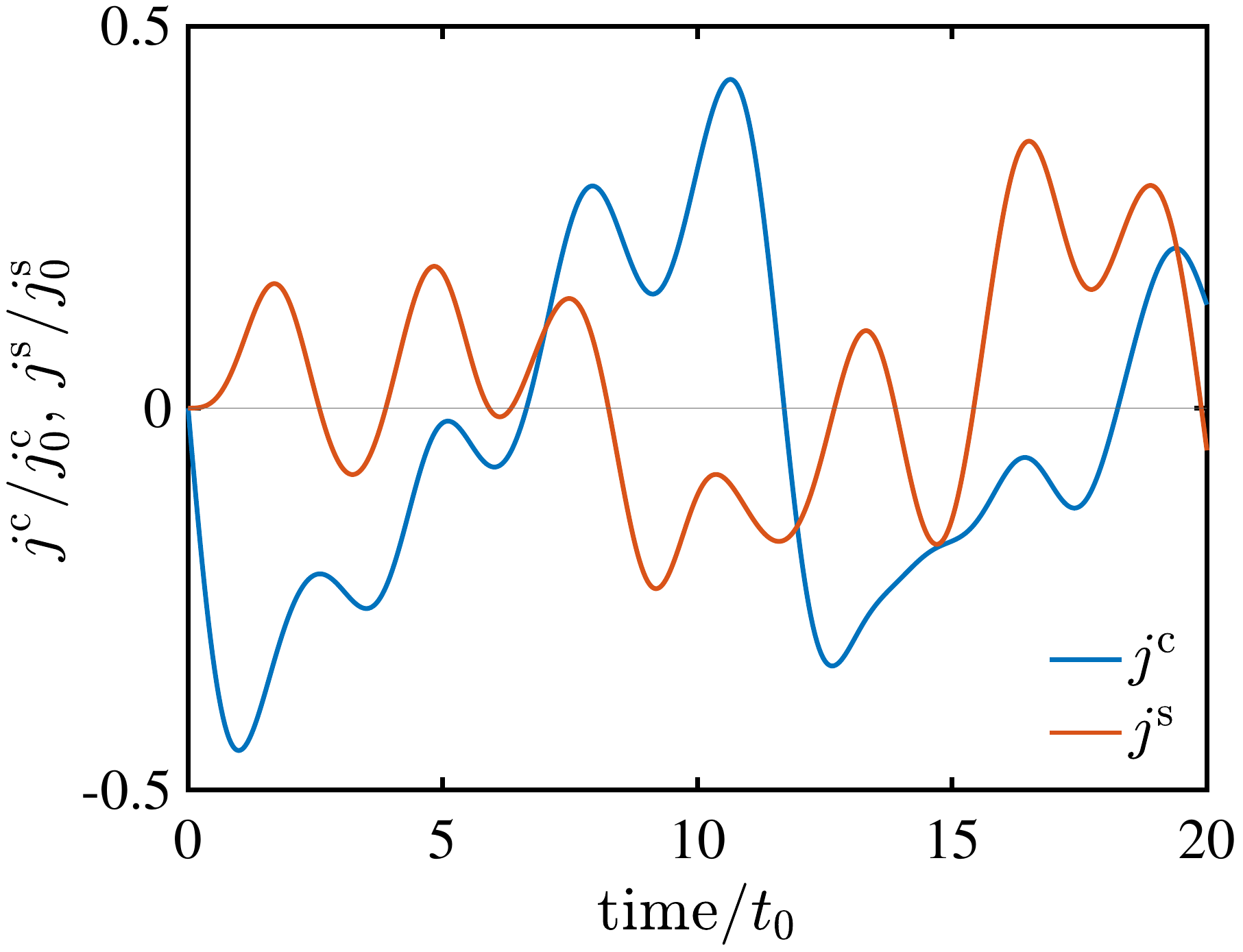}
\label{fig_jcjs_PTsym}}%
\quad
\subfigure[]{%
\includegraphics[bb=3 2 514 415,width=0.7\columnwidth]{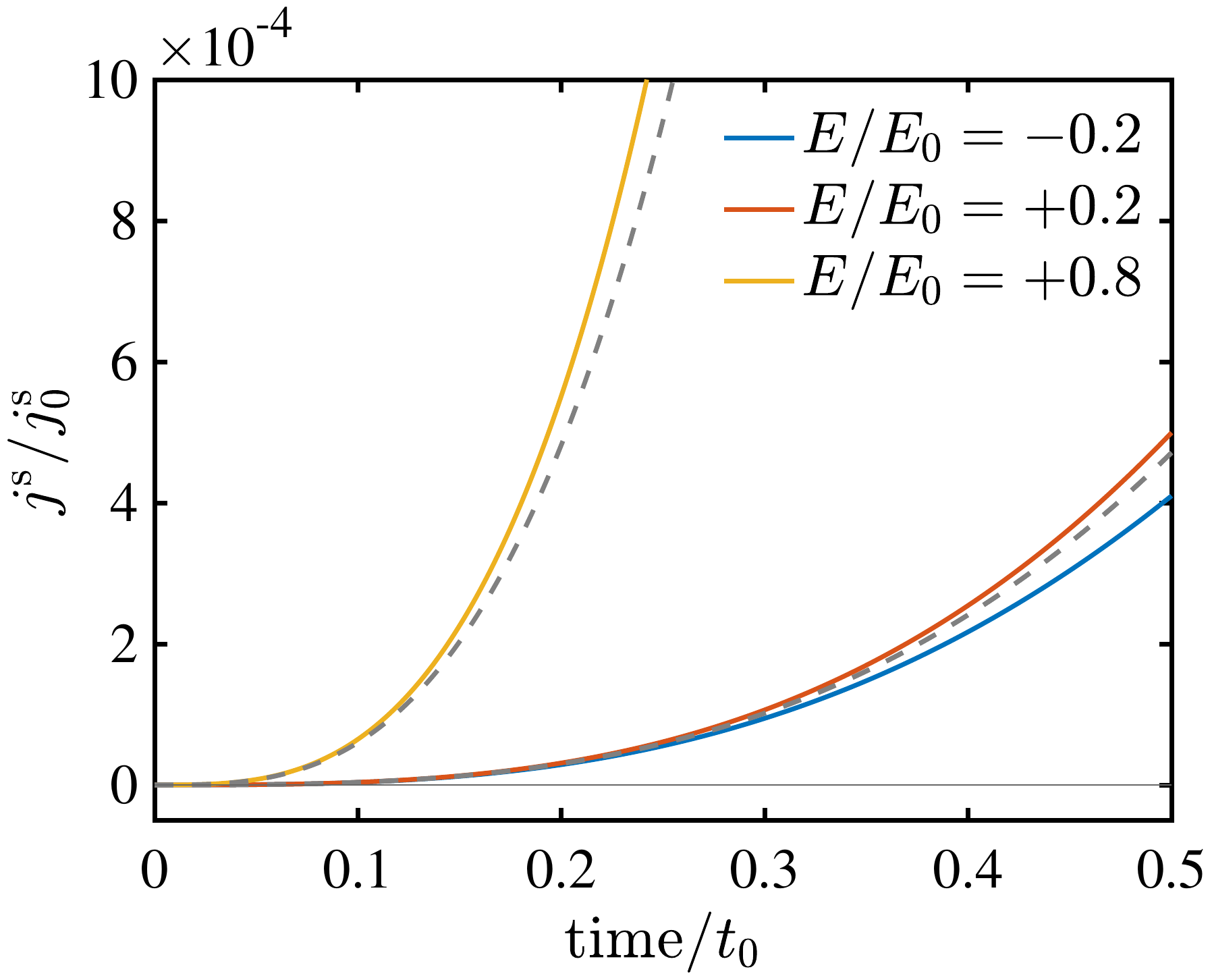}
\label{fig_js_dE_PTsym}}%
\caption{(Color online) 
Generalized model with PT symmetry and numerical results for tunneling charge/spin current from this model.
(a) Zigzag chain model with PT symmetry. 
The components in the model are the same as those in Fig.~\ref{fig_zigzagPandT},
except for the staggered magnetic order on each site (shown as blue circles $\odot, \otimes$).
(b) Time variations of charge and spin currents in the 
zigzag chain model with PT symmetry.
We used the parameters $(u/{w}, v/{w}, \phi_s/{w}, E/E_0)=(0.7, 0.4, 0.2, 0.8)$.
(c) Electric field dependence of the spin current shortly after the application of an electric field.
The dashed curves represent $j^{\mathrm{s}}(E, t)$ from the analytical formula (\ref{eq_approx_js_PTsym}).
We used the parameters $(u/{w}, v/{w}, \phi_s/{w})=(0.7, 0.4, 0.2)$.
}
\label{3figs_PTsym}
\end{figure*}

We now generalize the zigzag chain model in Sec.~\ref{zigzag_PandT}
by introducing staggered magnetic order on each site.
In this model, illustrated in Fig.~\ref{fig_zigzag_chain_PTsym},
inversion and time-reversal symmetries are both broken, but PT symmetry is preserved.
Introducing the magnetic order may seem to be artificial,
but it is known that for a zigzag chain model, which has itinerant electrons and localized spins, 
the staggered antiferromagnetic order along the chain is stabilized near half-filling \cite{Hayami2015}.

Now, the molecular field from the magnetic order is expressed as 
\begin{equation}
 \hat{\mathcal{H}}_{\mathrm{AF}} = \sum_{k\sigma} 
\hat{C}^{\dagger}_{k\sigma}
\begin{pmatrix}
 \phi_s \cdot \sigma & & & \\
 & -\phi_s \cdot \sigma & & \\
 & &\phi_s \cdot \sigma & \\
 & & & -\phi_s \cdot \sigma
\end{pmatrix}
\hat{C}_{k\sigma}
~.
\end{equation}
We add this term to the Hamiltonian (\ref{eq_Sigma_H(k)}) and derive the effective Hamiltonian
$H^{\sigma}(k)$ for $\Ket{s_{\sigma}}$ in the same way as in Sec.~\ref{zigzag_PandT}. Then, we obtain
\begin{equation}
H^{\sigma}(k)= \tilde{\bm{d}}^{\sigma}(k)\cdot\bm{\tau}\label{eq_Hamiltonian_PT} 
\end{equation}
with
\begin{equation}
 \tilde{\bm{d}}^{\sigma}(k) =({w}\cos (ka), v\sin (ka), \sigma [\phi_s - u\sin (ka)\cos (ka)])
\end{equation}
where $u, v, {w}$ are the same as those defined in Eq.~(\ref{eq_uvt_expression}).

In this model, the expression for the shift vector is rather complicated, and
we focus on whether a spin current can be generated from this model.
Figure~\ref{fig_jcjs_PTsym} plots charge and spin currents computed from the Hamiltonian (\ref{eq_Hamiltonian_PT}). 
The nonzero spin current shown here ensures that
tunneling spin current occurs even in insulators with spin degeneracy by the application of an electric field.
Again, oscillating behaviors of the charge and spin currents are evident, 
and we give the reason for this in Appendix \ref{appendix_oscillation}.

Let us focus on the time period shortly after the application of the electric field.
Then, the tunneling spin current from this model is approximated as
\begin{multline}
 j^{\mathrm{s}}(E,t)/j^{\mathrm{s}}_0\\
\simeq \frac{2}{3\pi} 
\left(\frac{1}{a^2{{w}}^2}\int_{\BZ}\dd k~ \egap^2{\Abs{A^{\uparrow}_{+-}}}^2R^{\uparrow}_{+-}\right)
\cdot \left(\frac{E}{E_0}\right)^2\left(\frac{t}{t_0}\right)^3\label{eq_approx_js_PTsym}
\end{multline}
in the PT symmetric model.
We derive this expression in Appendix \ref{sec_approx_tunneling_current}.
In Fig.~\ref{fig_js_dE_PTsym},
this formula fits the behavior of $j^{\mathrm{s}}(t)$ for each value of $E/E_0$.
We also find from Fig.~\ref{fig_js_dE_PTsym} that 
this spin current satisfies nonreciprocity i.e., 
the relation $j^{\mathrm{s}}(-E)\neq -j^{\mathrm{s}}(+E)$, 
which is expected from the lack of inversion symmetry in this model.

\section{\label{conclusions}Conclusions}

We investigated whether a tunneling spin current is induced by an electric field
in connection with the shift vector.
We focused on a system with PT symmetry but without spin rotational symmetry,
including a system with both inversion and time-reversal symmetries.
First, we find from the analytic calculations for a two-band model in Sec.~\ref{sec_formulation} 
that a shift vector is necessary for spin current generation.
Then, in Sec.~\ref{zigzag_PandT}, 
we demonstrate a model that has a shift vector and calculated the spin current numerically.
We confirm that even in a system with inversion and time-reversal symmetries,
a spin current can be generated because of the shift vector.
We extend this discussion to the general PT symmetric system, where
tunneling spin current also arises, in Sec.~\ref{zigzag_PTsym}.

Taking all this into account,
we can conclude that it is the shift vector, a purely geometric quantity, 
that causes a tunneling spin current in a system with spin degeneracy.
So far, we have only considered models where
a certain spin component is a good quantum number. However,
even in systems that do not satisfy this condition, 
we only need to consider 
doubly degenerate levels that exchange with each other under PT transformation,
rather than degenerate bands with opposite spins.
Our result can then be applied to this system, but its formulation is left for a future study.

The results of this study should help the search for materials useful in spintronics.
Even for highly symmetric insulators, which are considered less attractive for spintronics, 
we find that it is usually possible to generate tunneling spin currents.
Our results also show  
that materials with a relatively large tunneling spin current can be identified 
by estimating the shift vector for the material.
Although we have analyzed zigzag chains as an example, 
we expect that the shift vector also remains finite in many other crystal structures.

In addition, this study may lead to the proposal of a new spin current generation mechanism.
It is known that tunneling, as a response to a strong dc electric field,
undergoes a crossover in response to a weak ac electric field 
in the plane of the field strength and frequency \cite[Sec.~IIIA]{Aoki2014}. 
This study should stimulate 
theoretical research into the ac spin current response in PT symmetric systems
that corresponds to the tunneling spin current in this paper,
since it is easier to experimentally study an ac spin current response
than a dc response, which requires a strong electric field.

Finally, this study for a system with both inversion and time-reversal symmetries
should provide insight for the spin transport properties of other systems that 
have only inversion or time-reversal symmetry.
For example, this study may help us to understand 
the spin transport properties of chiral molecules where inversion symmetry is broken.
In some chiral molecules, highly spin-dependent transport called chirality-induced spin selectivity (CISS)
has recently been found \cite{Gohler2011a,Xie2011}, but its mechanism is still under debate \cite{Naaman2019a}.
As well as CISS, this study should be useful
for characterizing spin transport in a wide range of materials.

\begin{acknowledgments}
The author is grateful to Y. Kato, T. Morimoto, S. Kitamura and T. Oka for constructive discussions on the subject. 
The author wishes to thank E. Saito, D. Hirobe, H. Yamamoto, Y. Togawa and J. Kishine, for their helpful contributions.
This work is supported by the World-leading Innovative Graduate Study Program for 
Materials Research, Industry, and Technology (MERIT-WINGS) of the University of Tokyo.
\end{acknowledgments}

\appendix

\section{Expression for the tunneling current}
\label{appendix_expressions_current}
Here, we derive Eq.~(\ref{current_a_rep}) from the definition (\ref{eq_current_sum_bz}).
First, we find from the Schr\"{o}dinger equation (\ref{TDSE}) that 
\begin{align}
&\pdv{}{t}\Braket{\Psi^{\sigma}_{k_0}(t)|\hat{H}^{\sigma}(k_0(t))|\Psi^{\sigma}_{k_0}(t)} \nonumber\\
&= \Braket{\Psi^{\sigma}_{k_0}(t)|\hat{H}^{\sigma}(k_0(t))|\partial_t\Psi^{\sigma}_{k_0}(t)} + \mathrm{c.c.} \nonumber\\
&\qquad + \Braket{\Psi^{\sigma}_{k_0}(t)|\pdv{\hat{H}^{\sigma}(k_0(t))}{t}|\Psi^{\sigma}_{k_0}(t)} \\
&= \frac{1}{i\hbar}\Braket{\Psi^{\sigma}_{k_0}(t)|\left[\hat{H}(k_0(t))\right]^2|\Psi^{\sigma}_{k_0}(t)} 
+ \mathrm{c.c.}\nonumber\\
& \qquad + \pdv{k_0(t)}{t}\cdot \Braket{\Psi^{\sigma}_{k_0}(t)|
\left.\pdv{\hat{H}^{\sigma}(k)}{k}\right|_{k=k_0(t)}|\Psi^{\sigma}_{k_0}(t)}\\
& = -eE \Braket{\Psi^{\sigma}_{k_0}(t)|\hat{v}^{\sigma}_{k_0}(t)|\Psi^{\sigma}_{k_0}(t)}\label{eq_time_diff_Leibniz}
\end{align}
where $\hat{v}^{\sigma}_{k_0}(t)$ is the velocity operator
introduced in Sec.~\ref{subsec:P_j_def}.

Then, Eq.~(\ref{eq_a_expansion_of_Psi}) leads to
\begin{align}
&\Braket{\Psi^{\sigma}_{k_0}(t)|\hat{H}^{\sigma}(k_0(t))|\Psi^{\sigma}_{k_0}(t)}\nonumber\\
&= \epsilon_+(k_0(t)) \Abs{a^{\sigma}_{+k_0}(t)}^2 + \epsilon_-(k_0(t)) \Abs{a^{\sigma}_{-k_0}(t)}^2\\
&= \egap (k_0(t)) P^{\sigma}_{k_0}(t) + \epsilon_-(k_0(t))~,
\label{eq_psi_expectation}  
\end{align}
using the conservation of probability $\Braket{\Psi^{\sigma}_{k_0}(t)|\Psi^{\sigma}_{k_0}(t)} = 
\Abs{a^{\sigma}_{+k_0}(t)}^2+\Abs{a^{\sigma}_{-k_0}(t)}^2 = 1$.
Finally, combining Eq.~(\ref{eq_time_diff_Leibniz}) and Eq.~(\ref{eq_psi_expectation}), we have
\begin{multline}
-e\Braket{\Psi^{\sigma}_{k_0}(t)|\hat{v}^{\sigma}_{k_0}(t)|\Psi^{\sigma}_{k_0}(t)}\\
= \frac{1}{E}\pdv{}{t}\left[ \egap (k_0(t)) P^{\sigma}_{k_0}(t) \right]
+ \frac{1}{E}\pdv{\epsilon_-(k_0(t))}{t}~.
\label{eq_group_velocity}
\end{multline}
We then integrate both sides of Eq.~(\ref{eq_group_velocity})
over the BZ.
The second term in the right-hand side of Eq.~(\ref{eq_group_velocity}) vanishes and
we obtain Eq.~(\ref{current_a_rep}).

\section{Symmetry considerations of the Berry connection and shift vector}
\label{appendix_symmetry}
In this appendix, we consider the symmetry of the Berry connection and shift vector
reflecting the discrete symmetry of the spinful system.
Note that the case when the system has spin rotational symmetry is considered in \cite{Kitamura2020a}.
Let us focus on a system that is invariant under a discrete transformation, such as space inversion $\hat{\Pi}$,
time-reversal $\hat{\Theta}$ or PT transformation $\hat{\Pi}\hat{\Theta}$.
The Hamiltonian of this system satisfies
\begin{equation}
 \hat{{O}}\hat{H}(k)=\hat{H}(\chi\cdot k) \hat{{O}}\label{Commu_HU} 
\end{equation}
with $(\hat{{O}},\chi) = (\hat{\Pi}, -1), (\hat{\Theta}, -1), (\hat{\Pi}\hat{\Theta}, +1)$.

The eigenstates of the Hamiltonian are introduced as 
$\hat{H}(k)\ket{u_{\alpha}(k)}=\epsilon_{\alpha}(k)\ket{u_{\alpha}(k)}$
where $\alpha = (n,\sigma)$ is a combination of band and spin indices.
We then find from Eq.~(\ref{Commu_HU}) that 
$\hat{{O}}\ket{u_{\alpha}(k)}$ is one of the eigenstates of $\hat{H}(\chi\cdot k)$ 
with energy $\epsilon_{\alpha}(k)$.
We consider this state labeled with another band index $\tilde{\alpha}$ such that
\begin{equation}
\ket{u_{\tilde{\alpha}}(\chi\cdot k)} = e^{-i\theta_{\alpha}(k)}\hat{{O}}\ket{u_{\alpha}(k)},\quad
\epsilon_{\tilde{\alpha}}(\chi\cdot k)= \epsilon_{\alpha}(k)\label{utilde_def}
\end{equation}
where the difference in the phase $\theta_{\alpha}(k)$ remains in general.

From the definition of the Berry connection and Eq.~(\ref{utilde_def}), we find that
\begin{align}
&A_{\tilde{\alpha}\tilde{\beta}}(\chi\cdot k)
= \Braket{u_{\tilde{\alpha}}(\chi\cdot k)|i\partial_{\chi\cdot k}|u_{\tilde{\beta}}(\chi\cdot k)} \\
&= \chi \cdot 
\left[\Braket{\hat{{O}}u_{\alpha}(k)|\hat{{O}}i\partial_{k}u_{\beta}(k)}\right.\nonumber\\
&+\left.\left(\partial_k\theta_{\beta}\right)
\Braket{\hat{{O}}u_{\alpha}(k)|\hat{{O}}u_{\beta}(k)}\right]
\cdot e^{i(\theta_{\alpha}(k)-\theta_{\beta}(k))}~.\label{eq_tochu}
\end{align}
Here, $\hat{O}=\hat{\Pi}$ is a unitary operator and $\hat{O}=\hat{\Theta}, \hat{\Pi}\hat{\Theta}$ are antiunitary operators.
These operators obey
\begin{equation}
 \Braket{\hat{O} \psi|\hat{O} \phi}
= 
\begin{cases}
\Braket{\psi|\phi} & \text{if $\hat{O}$ is unitary ($\mathrm{U}$)}\\    
\Braket{\phi|\psi} & \text{if $\hat{O}$ is antiunitary ($\mathrm{A}$)} 
\end{cases}\label{UAU}
\end{equation}
for arbitrary states $\ket{\psi}, \ket{\phi}$.
Hereafter, we substitute $(\mathrm{U}), (\mathrm{A})$ for unitary or antiunitary
$\hat{O}$, respectively.

Then, we find from Eq.~(\ref{eq_tochu}) that
\begin{align}
&A_{\tilde{\alpha}\tilde{\beta}}(\chi\cdot k)\nonumber\\
&= 
\begin{cases}
\chi \cdot \left[
+A_{\alpha\beta}(k) +(\partial_k\theta_{\beta})\delta_{\alpha\beta}
\right]e^{i(\theta_{\alpha}(k)-\theta_{\beta}(k))}
& (\mathrm{U}) \\
\chi \cdot \left[
-A^{\ast}_{\alpha\beta}(k) +(\partial_k\theta_{\beta})\delta_{\alpha\beta}
\right]e^{i(\theta_{\alpha}(k)-\theta_{\beta}(k))}
& (\mathrm{A} )
\end{cases}
~.\label{eq_Asym}
\end{align}

Furthermore, the shift vector $R_{\tilde{\alpha}\tilde{\beta}}(\chi\cdot k)$  ($\alpha\neq \beta$) satisfies
\begin{align}
& R_{\tilde{\alpha}\tilde{\beta}}(\chi\cdot k)\nonumber\\
&= A_{\tilde{\alpha}\tilde{\alpha}}(\chi\cdot k) -A_{\tilde{\beta}\tilde{\beta}}(\chi\cdot k) 
- \partial_{\chi\cdot k}\arg A_{\tilde{\alpha}\tilde{\beta}}(\chi\cdot k) \\
&= 
\begin{cases}
+\chi\cdot \left[
A_{\alpha\alpha}(k) -A_{\beta\beta}(k) -\partial_k\arg A_{\alpha\beta}(k)
\right]
 & (\mathrm{U})\\   
-\chi\cdot \left[
A_{\alpha\alpha}(k) -A_{\beta\beta}(k) -\partial_k\arg A_{\alpha\beta}(k)
\right]
& (\mathrm{A})
\end{cases}
~.\label{eq_Rsym}
\end{align}
where $\theta_{\alpha,\beta}$ vanishes because of the gauge invariance of the shift vector.

From Eqs.~(\ref{eq_Asym}) and (\ref{eq_Rsym}), we find the following relations: 
\begin{subequations}
\label{THEOREM}
\begin{gather}
 \Abs{A_{\tilde{\alpha}\tilde{\beta}}(\chi\cdot k)} = \Abs{A_{\alpha\beta}(k)}\qquad (\alpha\neq \beta),\\
R_{\tilde{\alpha}\tilde{\beta}}(\chi\cdot k) = 
\begin{cases}
 +\chi\cdot R_{\alpha\beta}(k) & (\mathrm{U})\\
 -\chi\cdot R_{\alpha\beta}(k) & (\mathrm{A})
\end{cases}
~.
\end{gather}
\end{subequations}

Let us now consider a PT symmetric system.
In this case, $\hat{O}=\hat{\Pi}\hat{\Theta}$ is antiunitary $(\mathrm{A})$ and $\chi = +1$.
If we take $\alpha,\beta = (n,\uparrow), (m,\uparrow)$,
we then find $\tilde{\alpha}, \tilde{\beta}=(n,\downarrow), (m,\downarrow)$.
Taking these into account, we immediately obtain Eqs.~(\ref{ARRelationwPTsym}) from Eqs.~(\ref{THEOREM}).

Similarly, in a system with both inversion and time-reversal symmetries, 
we obtain
\begin{subequations}
\begin{equation}
 \Abs{A^{\sigma}_{nm}(-k)} =\Abs{A^{\sigma}_{nm}(k)},\quad
 R^{\sigma}_{nm}(-k)= -R^{\sigma}_{nm}(k)~,
\label{eq_inversion_AR}
\end{equation}
from the inversion symmetry, and
 \begin{equation}
\Abs{A^{\uparrow}_{nm}(-k)} =\Abs{A^{\downarrow}_{nm}(k)},\quad
 R^{\uparrow}_{nm}(-k)= +R^{\downarrow}_{nm}(k)   
\label{eq_time_reversal_AR}
 \end{equation}
\end{subequations}
from the time-reversal symmetry. 
Combining Eqs.~(\ref{eq_inversion_AR}) and (\ref{eq_time_reversal_AR}), we get
Eqs.~(\ref{ARRelationwPTsym}) and (\ref{ARRelationwPandTsym}).

\section{Reason for tunneling current oscillation}
\label{appendix_oscillation}
We consider the components of the Hamiltonian of the zigzag chain models
shown in Eqs.~(\ref{eq_Hamiltonian_PandT}) and (\ref{eq_Hamiltonian_PT}).
We find that in both models, $\hat{H}^{\sigma}{(k_0(t))}$ is time-periodic with period $T=2\pi \hbar / (e|E|a)$. 
In this case, Floquet theory \cite[Sec.~3.6]{Teschl2012} gives some conditions for $\Ket{\Psi^{\sigma}_{k_0}(t)}$ 
as a solution of Eq.~(\ref{TDSE}).
Note that we also have to pay attention to conservation of probability
$\Braket{\Psi^{\sigma}_{k_0}(t)|\Psi^{\sigma}_{k_0}(t)} =1$.
Then, we obtain the expression 
\begin{equation}
 \Ket{\Psi^{\sigma}_{k_0}(t)} = \sum_{\nu=1,2}b^{\sigma}_{\nu k_0}
e^{-i\varepsilon^{\sigma}_{\nu}t/\hbar}\Ket{\Phi^{\sigma}_{\nu k_0}(t)}.\label{eq_Floquet_sol} 
\end{equation}
Here, $\Ket{\Phi^{\sigma}_{\nu k_0}(t)}$ is a time-periodic state with period $T$, 
and the Floquet exponent $i\varepsilon^{\sigma}_{\nu}$ is imaginary.
We also introduced the coefficients $b^{\sigma}_{\nu k_0}$.
Note that $\varepsilon^{\sigma}_{\nu}$ is independent of $k_0$ 
because Eq.~(\ref{TDSE}) shows that 
the difference in $k_0$ is reduced to the shift of the origin of time, 
which has no effect on the Floquet exponent of the solution.

Furthermore, when the system has time-reversal symmetry, 
$\varepsilon^{\sigma}_{\nu}$ is also independent of $\sigma$
because 
the time-reversal operation (antiunitary) and the subsequent change of variable $t\to -t$ converts
the Schr\"{o}dinger equation (\ref{TDSE}) and the solution (\ref{eq_Floquet_sol})
with the label $(k_0,\sigma)$ into 
those with $(-k_0, -\sigma)$, but the Floquet exponent is unchanged.

Now, we find
 \begin{align}
&\Braket{\Psi^{\sigma}_{k_0}(t)|\hat{H}^{\sigma}(k_0(t))|\Psi^{\sigma}_{k_0}(t)}\nonumber\\
&= \sum_{\mu, \nu}{b^{\sigma}_{\mu k_0}}^{\ast}b^{\sigma}_{\nu k_0}
e^{i(\varepsilon^{\sigma}_{\mu}-\varepsilon^{\sigma}_{\nu})t/\hbar}
\Braket{\Phi^{\sigma}_{\mu k_0}(t)|\hat{H}^{\sigma}(k_0(t))|\Phi^{\sigma}_{\nu k_0}(t)}\nonumber\\
  &=  \mathcal{P}_0(t)+\mathcal{P}_1(t)\cos (\omega^{\sigma}_{12}t) + \mathcal{P}_2(t)\sin (\omega^{\sigma}_{12}t)
\end{align}
where $\mathcal{P}_{0,1,2}(t)$ are all real periodic function with period $T$,
and $\omega^{\sigma}_{12}\equiv \abs{\varepsilon^{\sigma}_1 -\varepsilon^{\sigma}_2}/\hbar$.
Thus, we get from Eq.~(\ref{eq_current_sum_bz}) and Eq.~(\ref{eq_time_diff_Leibniz}),
\begin{align}
 j^{\sigma}&= 
\frac{1}{EL}
\sum_{k_0\in \BZ} 
\pdv{}{t}\Braket{\Psi^{\sigma}_{k_0}(t)|\hat{H}^{\sigma}(k_0(t))|\Psi^{\sigma}_{k_0}(t)}\\
&= \mathcal{Q}_0(t)+{\mathcal{Q}}_1(t)\cos (\omega^{\sigma}_{12}t) + {\mathcal{Q}}_2(t)\sin (\omega^{\sigma}_{12}t) 
\label{eq_osc_current_rep_freq}
\end{align}
where $\mathcal{Q}_{0,1,2}(t)$ are all periodic with period $T$ and 
$j^{\sigma}$ has no dc component.
As we mentioned above, $\omega^{\uparrow}_{12}=\omega^{\downarrow}_{12}$ also holds 
in time-reversal symmetric systems.

\section{Approximation of tunneling current shortly after application of electric field}
\label{sec_approx_tunneling_current}
In the adiabatic limit, Eq.~(\ref{eq_Tprod}) leads to
\begin{multline}
\begin{pmatrix}
 a^{\sigma}_{+k_0}(t)e^{-i\arg A^{\sigma}_{+-}(k_0)}\\
 a^{\sigma}_{-k_0}(t)
\end{pmatrix}
\simeq \left[\openone +i\int_{k_0}^{k_0(t)}
\abs{A^{\sigma}_{+-}(k)} \right. \\
\left.\cdot 
\begin{pmatrix}
 0 & e^{-i{\Delta^{\sigma}_{+-}(k,k_0)}}\\
e^{+i{\Delta^{\sigma}_{+-}(k,k_0)}} & 0
\end{pmatrix}
\dd k \right]
\begin{pmatrix}
 0\\  1
\end{pmatrix} 
~. 
\end{multline}
For convenience, we write $A_0\equiv \Abs{A^{\sigma}_{+-}}$, $R_0\equiv R^{\uparrow}_{+-}$,
$q\equiv -eEt/\hbar$ and $\lambda \equiv (\egap /eE ) + \sigma\cdot R_0$.
We assume that $t$ is small enough to satisfy $\Abs{q}a\ll 1$ where $a$ is the lattice constant, hereafter taken as $a=1$.
We also consider the band parameter ${w}$ shown in Eq.~(\ref{eq_uvt_expression}) as a typical energy scale, and
set ${w}=1$ for simplicity.
By using the Taylor expansion with $q$, we find that
\begin{align}
& P^{\sigma}_{k_0}(E, t)= \Abs{a^{\sigma}_{+k_0}}^2\\
&=\Abs{ \int_{k_0}^{k_0+q}\dd k~ {A_0}\exp\left(-i\int_{k_0}^{k}\dd k'~\lambda\right) }^2\\
&=\left| A_0(k_0)\frac{q^1}{1!}
+\left.(\partial_k -i\lambda){A_0}\right|_{k=k_0}\frac{q^2}{2!}\right.\nonumber\\
&\left. +\left.(\partial_k -i\lambda)^2{A_0}\right|_{k=k_0}\frac{q^3}{3!}
+\left.(\partial_k -i\lambda)^3{A_0}\right|_{k=k_0}\frac{q^4}{4!}
+\cdots\right|^2\\
&\equiv P^{\sigma}_{k_0}(E, q)~.\label{eq_P_Taylor}
\end{align}

The tunneling current Eq.~(\ref{current_a_rep}) is expressed as
\begin{equation}
j^{\sigma} = \frac{-e}{\hbar}\int_{\BZ}\frac{\dd k_0}{2\pi}~
\frac{\partial}{\partial q}
\left[\egap (k_0+q) P^{\sigma}_{k_0}(E,q)
\right]~.
\end{equation}
We expand this formula with $q$ by using 
Eq.~(\ref{eq_P_Taylor}) and 
the expansion
$\egap (k_0+q) = \egap(k_0) + \left.\partial_{k}\egap\right|_{k=k_0} q^1/1! + \cdots$.
Then, we find
\begin{multline}
 \egap (k_0+q) P^{\sigma}_{k_0}(E,q)/q^2 = c_{0,0}+c_{1,0}q^1\\
+\left[c_{2,2}(E/E_0)^{-2} + c_{2,1}(E/E_0)^{-1}\sigma + c_{2,0}\right]q^2\\
+\left[c_{3,2}(E/E_0)^{-2} + c_{3,1}(E/E_0)^{-1}\sigma + c_{3,0}\right]q^3\\
+ \mathcal{O}\left((t/t_0)^4\right)
\label{eq_egapP_expand} 
\end{multline}
where we introduced $E_0= {w}/(ea)$ and $t_0= \hbar/{w}$. 
The coefficients $c_{l,m}$ consist of $\egap$, $A_0$ and $R_0$, for example,
\begin{gather}
 c_{0,0} = \left.\egap {A_0}^2\right|_{k=k_0}, \quad c_{2,1}= -\frac{1}{6}\left.{\egap}^2{A_0}^2R_0\right|_{k=k_0},\\
c_{3,1} = \frac{1}{12}\left.\left[
\egap(\partial_{k}\egap){A_0}^2R_0 - \partial_{k}({\egap}^2{A_0}^2R_0)
\right]\right|_{k=k_0}~.
\end{gather}
Then, the leading term of the charge current $j^{\mathrm{c}} = j^{\uparrow} + j^{\downarrow}$ is given as follows: 
\begin{align}
& j^{\mathrm{c}}(E, t)
\simeq 2\cdot \frac{-e}{\hbar}
 \int_{\BZ}\frac{\dd k_0}{2\pi} c_{0,0}\cdot 2q^1\\
 &= -\frac{2}{\pi}j^{\mathrm{c}}_0\cdot \left(
\frac{1}{a{w}}\int_{\BZ}\dd k~ \egap {A_0}^2
\right) \cdot \left(\frac{E}{E_0}\right)^1\left(\frac{t}{t_0}\right)^1~.
\label{eq_approx_jc}
\end{align}
Here, we reinserted the powers of $a$ and $w$ from dimensional analysis.
The leading term of the spin current $j^{\mathrm{s}}= (\hbar/2(-e))\cdot (j^{\uparrow}-j^{\downarrow})$ 
in a PT symmetric system is also given as follows:
\begin{equation}
 j^{\mathrm{s}}(E, t)
\simeq  2\cdot \frac{\hbar}{2(-e)}\cdot \frac{-e}{\hbar}\int_{\BZ}\frac{\dd k_0}{2\pi}
c_{2,1}\left(\frac{E}{E_0}\right)^{-1}\cdot 4q^3~.\label{eq_before_approx_js_PTsym}
\end{equation}
This equation leads to Eq.~(\ref{eq_approx_js_PTsym}).

However, in a system with both inversion and time-reversal symmetries, 
the integral in Eq.~(\ref{eq_before_approx_js_PTsym}) vanishes 
because the coefficient $c_{2,1}(k)$ is an odd function of $k$.
In general, $c_{l,m}(-k_0) = (-1)^{l+m}c_{l,m}(k_0)$ holds
because the expansion (\ref{eq_egapP_expand}) is valid
regardless of the inversion transformation
$(E, \sigma, q)\to (-E, +\sigma, -q)$.
Therefore, the leading term of $j^{\mathrm{s}}$ in this system is given as
\begin{equation}
 j^{\mathrm{s}}(E, t)
 \simeq  2\cdot \frac{\hbar}{2(-e)}\cdot \frac{-e}{\hbar}\int_{\BZ}\frac{\dd k_0}{2\pi}
c_{3,1}
\cdot\left(\frac{E}{E_0}\right)^{-1} 5q^4,
\end{equation}
which leads to Eq.~(\ref{eq_approx_js_PandT}).



%

\onecolumngrid
\clearpage 

\renewcommand{\theequation}{S\arabic{equation}}
\renewcommand{\thefigure}{S\arabic{figure}}
\renewcommand{\thetable}{S\arabic{table}}
\renewcommand{\thesection}{S\arabic{section}}
\setcounter{equation}{0}
\setcounter{figure}{0}
\setcounter{table}{0}
\setcounter{page}{1}
\makeatletter

\begin{center}
\textbf{\large
Supplemental Material for \\
``Tunneling spin current in a system with spin degeneracy''
}
\end{center}
\vspace{20pt}

In this material, we provide additional numerical results 
for the zigzag chain models introduced in Sec.~\ref{zigzag_PandT} and Sec.~\ref{zigzag_PTsym} in the main text.
These results ensure the accuracy of our calculations and support the conclusions drawn in the main text.

The outline of the material is as follows. 
We first examine how accurately the conservation of probability during the interband transition is satisfied
in our computation.
We then show the Fourier spectrum of the oscillating tunneling current, 
and check that each peak in the spectrum is at the position expected from Appendix \ref{appendix_oscillation}.
Next, we consider how the parameter $v$ in the Hamiltonian (\ref{eq_Hamiltonian_PandT}) affects
the behavior of the spin current shortly after the application of an electric field.
Finally, we focus on
the electric field dependence of the charge current shortly after the application of the electric field.

\section*{\label{num_err_consev_prob}S1: Numerical accuracy regarding the conservation of probability}

In our models, the conservation of probability is described as follows: 
\begin{equation}
\frac{1}{N_\mathrm{ele}}\sum_{k_0\in \BZ}\sum_{\sigma = \uparrow, \downarrow}
\Braket{\Psi^{\sigma}_{k_0}(t)|\Psi^{\sigma}_{k_0}(t)} = 1~.\label{eq_conservation_prob} 
\end{equation}
Here, 
\begin{equation}
 N_{\mathrm{ele}} = 2\cdot L/\Abs{\UC} \label{eq_N_ele}
\end{equation}
is the total number of electrons that contribute to the tunneling process where
$\Abs{\UC} = 2a$ is the length of a unit cell and the factor two is 
due to the spin degree of freedom. 

Let us now compute the left-hand side of Eq.~(\ref{eq_conservation_prob})
and compare it to the right-hand side ($=1$)
to estimate any numerical errors in our computation.
We first take the tight-binding Hamiltonian 
shown in Eq.~(\ref{eq_Hamiltonian_PandT}) or Eq.~(\ref{eq_Hamiltonian_PT}),
and get $\Ket{\Psi^{\sigma}_{k_0}(t)}$ by solving the Schr\"{o}dinger equation (\ref{TDSE}) with the Runge--Kutta method.
Then, we take the sum in Eq.~(\ref{eq_conservation_prob}) for a sufficiently large $N_{\mathrm{ele}}$, 
and obtain the left-hand side of Eq.~(\ref{eq_conservation_prob}) at each time.

\begin{figure}[h]
\subfigure[]{%
\includegraphics[bb=3 2 517 415,width=0.45\columnwidth]{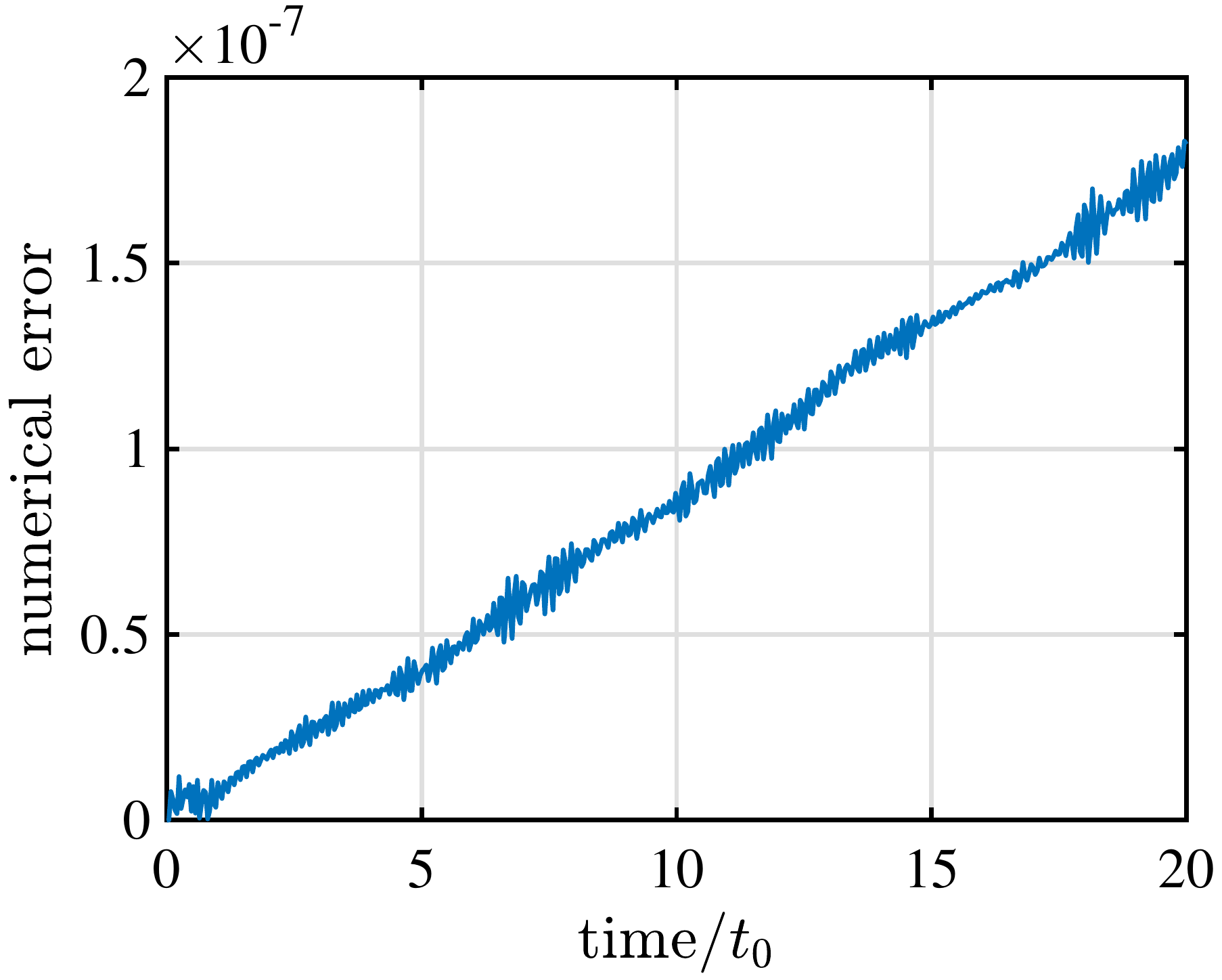}
\label{fig_num_err_PandT}
}%
\quad
\subfigure[]{%
\includegraphics[bb=3 2 517 415,width=0.45\columnwidth]{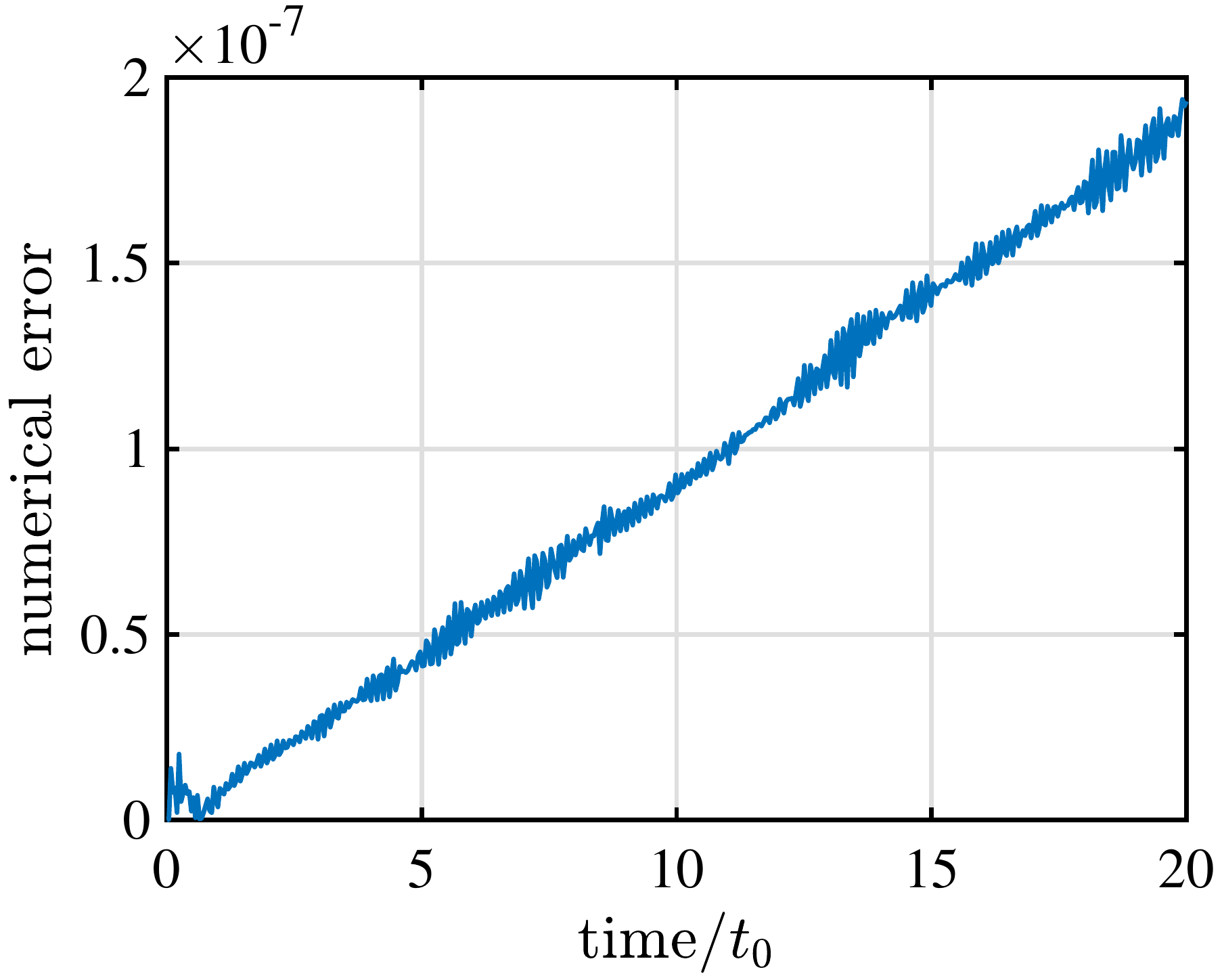}
\label{fig_num_err_PT}
}%
\caption{
Numerical errors can be seen in the difference between the two sides of Eq.~(\ref{eq_conservation_prob}).
In both figures, we set $N_{\mathrm{ele}}=200$.
(a) Error for the system with both inversion and time-reversal symmetries.
We used the same parameters $(u/{w}, v/{w}, E/E_0)=(0.7, 0.4, 0.8)$ as in Fig.~\ref{fig_charge_spin_current}.
(b) Error for the PT symmetric system.
We used the same parameters $(u/{w}, v/{w}, \phi_s/{w}, E/E_0)=(0.7, 0.4, 0.2, 0.8)$ 
as in Fig.~\ref{fig_jcjs_PTsym}.
}
\label{fig_num_error}
\end{figure}

Figure~\ref{fig_num_error} plots the difference between the two sides of Eq.~(\ref{eq_conservation_prob})
computed in this way.
The numerical error is found to be at most $3\times 10^{-7}$ in $0 < t/t_0 < 20$ from the figures.
We consider this error to be of the same order of magnitude as that in the expectation value of the tunneling current 
because the expectation value (\ref{eq_current_sum_bz}) is computed in the same way as 
the left-hand side of Eq.~(\ref{eq_conservation_prob}),
just as the sum (\ref{eq_current_sum_bz}) corresponds to the sum (\ref{eq_conservation_prob}).
Thus, the charge and spin current 
plotted in Fig.~\ref{fig_charge_spin_current} and Fig.~\ref{fig_jcjs_PTsym}
are considered to be correct up to order $10^{-7}$--$10^{-6}$. 
This estimate confirms the validity of 
the nonzero spin currents shown in these figures, which are not attributed to numerical errors but 
are generated from the zigzag chain models.

\section*{\label{sec:fft}S2: Fourier spectrum of oscillating current}
\begin{figure}[h]
\subfigure[]{%
\includegraphics[bb=3 3 521 382,width=0.7\columnwidth]{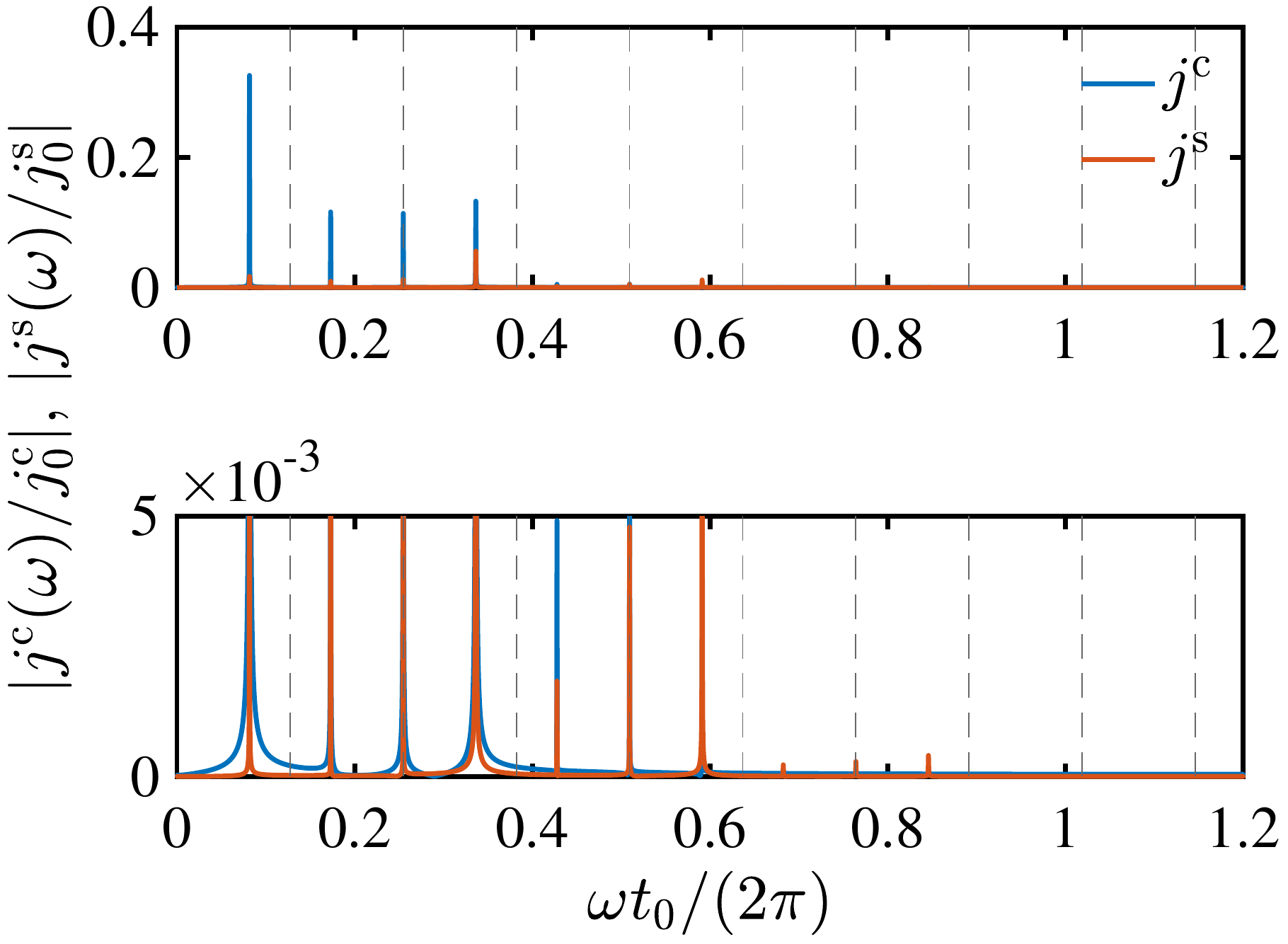}
\label{fig_fft_PandT}
}%
\\
\vspace{20pt}
\subfigure[]{%
\includegraphics[bb=3 3 521 375,width=0.7\columnwidth]{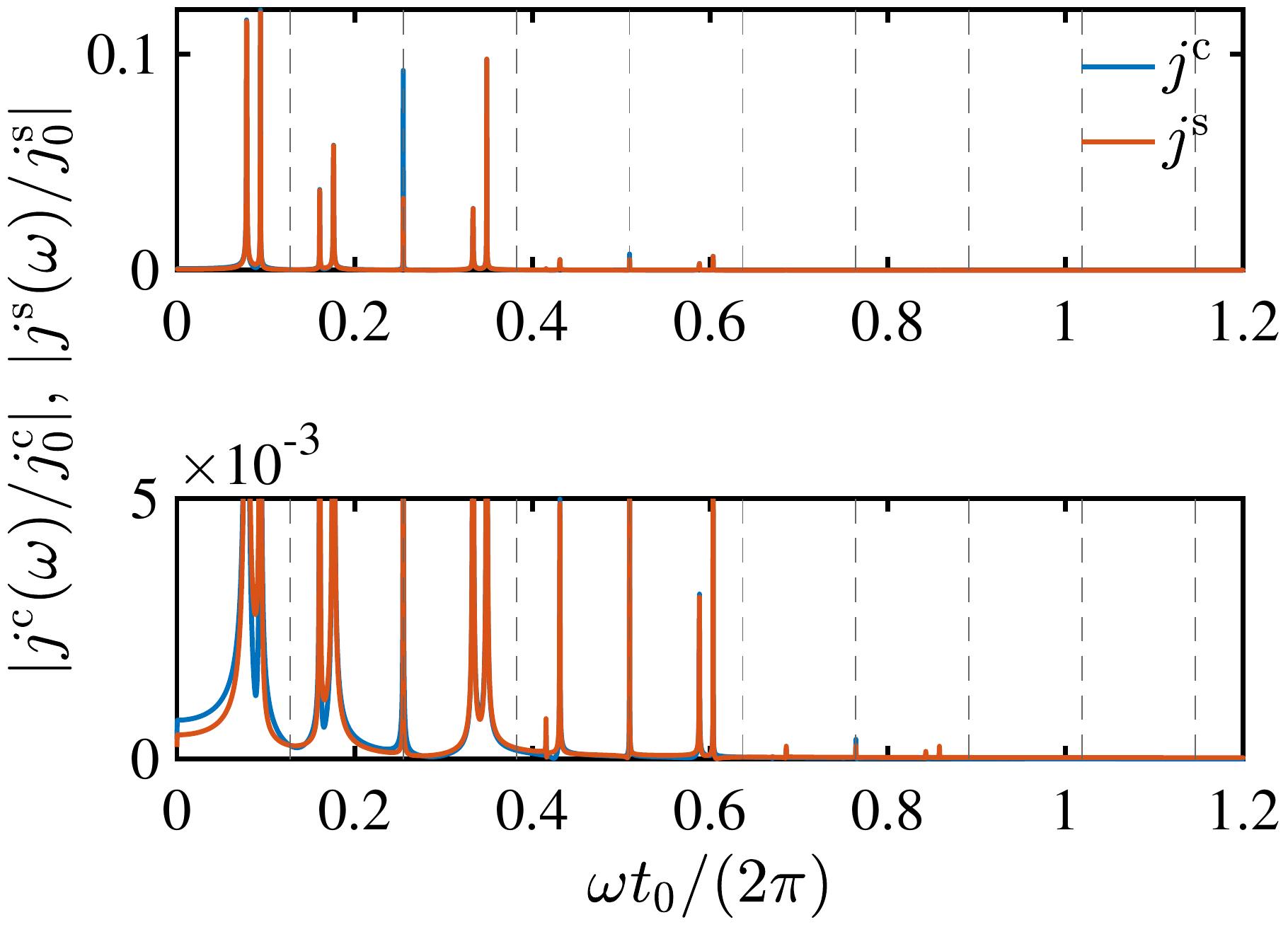}
\label{fig_fft_PTsym}
}%
\caption{(Color online) 
Fourier spectrum of charge and spin currents.
In both panels (a) and (b), the spectrum in the upper row is enlarged in the lower row.
The vertical dashed lines denote the frequencies $\omega t_0 = n\cdot E/E_0$ ($n=1, 2, \dots$).
(a) Model with both inversion and time-reversal symmetries. 
We used the parameters 
$(u/{w}, v/{w}, E/E_0)=(0.7, 0.4, 0.8)$. 
(b) PT symmetric model.
We used the parameters $(u/{w}, v/{w}, \phi_s/{w}, E/E_0)=(0.7, 0.4, 0.2, 0.8)$.
}
\label{fig_fft}
\end{figure}

In the main text, we observe oscillating behavior of the charge and spin currents $j^{\mathrm{c}}(t),j^{\mathrm{s}}(t)$
in Fig.~\ref{fig_charge_spin_current} and Fig.~\ref{fig_jcjs_PTsym},
explained in Appendix \ref{appendix_oscillation}.
To clarify this point, we derive 
the Fourier spectrum of the oscillating currents, shown in Fig.~\ref{fig_fft}.

We first focus on Fig.~S\ref{fig_fft_PandT},
which corresponds to the spectrum for the model with both inversion and time-reversal symmetries. 
We find that all the charge current peaks
overlap those of the spin current, and their positions can be expressed as 
\begin{equation}
 \omega t_0 = n\cdot E/E_0, \ n\cdot E/E_0 \pm 2\pi \delta \qquad (n=1, 2, \dots )
\label{eq_peaks_obs}
\end{equation}
where $\delta = 0.046$.
These peaks are predicted in Eq.~(\ref{eq_osc_current_rep_freq}), such that
\begin{equation}
 \omega t_0 = n\cdot E/E_0, \ n\cdot E/E_0 \pm \omega_{12}t_0 \qquad (n=1, 2, \dots )~.
\label{eq_peaks_theo}
\end{equation}
Here, we used the relation $(2\pi/T) t_0=(eEa/\hbar)\cdot (\hbar/{w})=E/E_0$
and introduced $\omega_{12}\equiv \omega^{\uparrow}_{12} = \omega^{\downarrow}_{12}$.
We find that Eqs.~(\ref{eq_peaks_obs}) and (\ref{eq_peaks_theo}) have the same form.
This agreement supports the analytical calculations in Appendix \ref{appendix_oscillation}.

We next focus on Fig.~S\ref{fig_fft_PTsym}, which corresponds to the spectrum for the PT symmetric model.
Again, we find that all the peaks of the charge current 
overlap those of the spin current, with their positions given by
\begin{equation}
\omega t_0 = n\cdot E/E_0, \ n\cdot E/E_0 \pm 2\pi \delta_1, \  n\cdot E/E_0 \pm 2\pi \delta_2 \qquad (n=1, 2, \dots ) 
\label{eq_peaks_obs_PT}
\end{equation}
where $\delta_1 = 0.033$, $\delta_2 = 0.049$.
These peaks are predicted in Eq.~(\ref{eq_osc_current_rep_freq}) such that
\begin{equation}
\omega t_0 = n\cdot E/E_0, \ n\cdot E/E_0 \pm \omega^{\uparrow}_{12}t_0 , \ 
n\cdot E/E_0 \pm \omega^{\downarrow}_{12}t_0 
\qquad (n=1, 2, \dots )~.
\label{eq_peaks_theo_PT}
\end{equation}
We now find that Eqs.~(\ref{eq_peaks_obs_PT}) and (\ref{eq_peaks_theo_PT}) have the same form.
This agreement also supports the analytical calculations in Appendix \ref{appendix_oscillation}.

\section*{\label{sec:js_v}S3: Spin current dependence on $v$}
In Sec.~\ref{zigzag_PandT}, we focus on 
a model with both inversion and time-reversal symmetries
where the shift vector is explicitly given by Eq.~(\ref{eq_Rrep}) and is proportional to the product $uv{w}$.
Note that $v\neq 0$ originates from the breaking of reflection symmetry in the model.

\begin{figure}[h]
\includegraphics[bb=3 2 514 415,width=0.45\columnwidth]{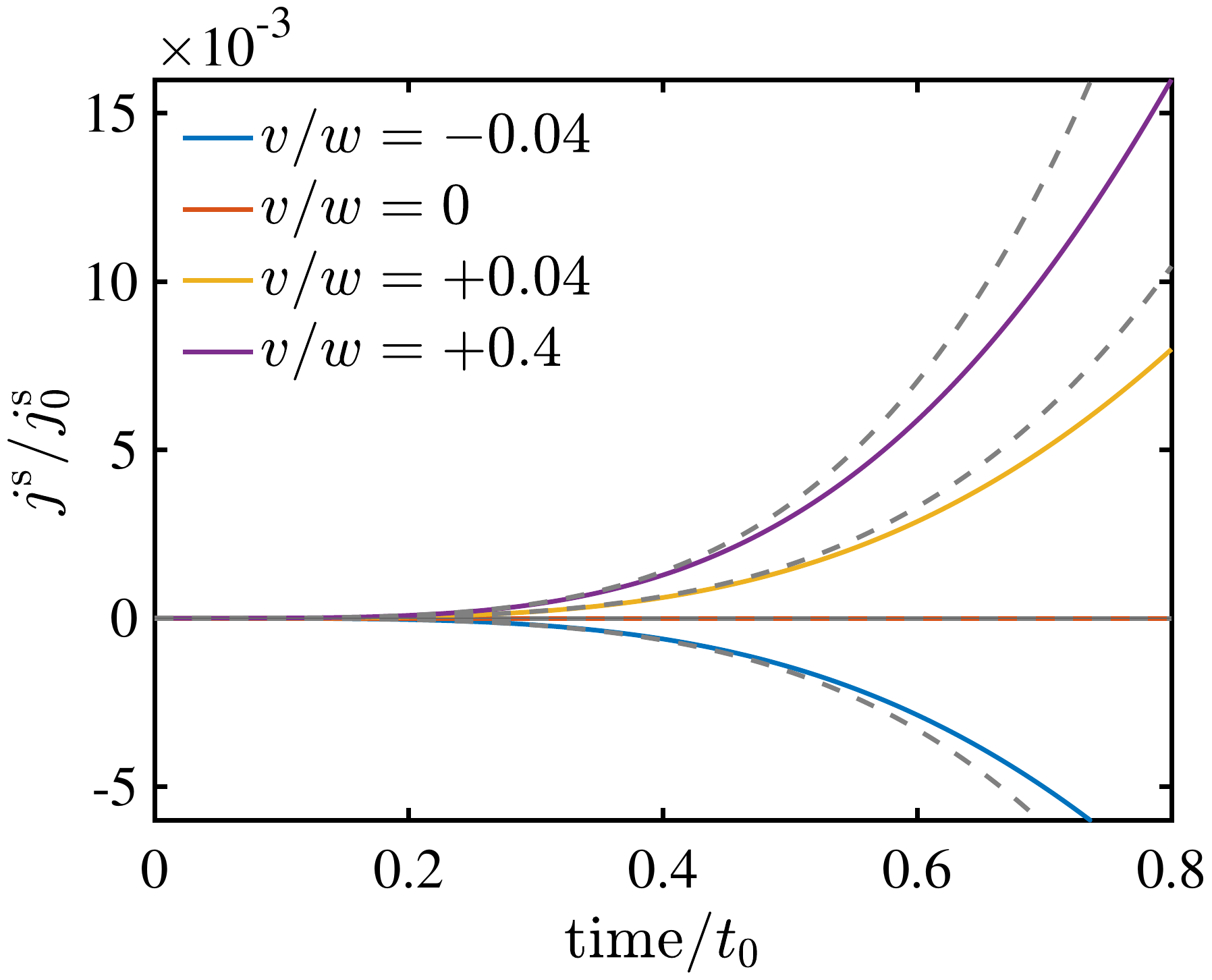}
\caption{(Color online) 
Time variation of spin current 
shortly after the application of an electric field for different values of $v$.
The dashed curves represent $j^{\mathrm{s}}(E, t)$ from the  analytical formula (\ref{eq_approx_js_PandT}).
We used the parameters $(u/{w}, E/E_0)=(0.7, 0.8)$.
} 
\label{fig_js_dv_PandT}
\end{figure}

Figure~\ref{fig_js_dv_PandT} plots the spin current dependence on $v$ in this system.
We find that the analytical formula (\ref{eq_approx_js_PandT})
fits the behavior of $j^{\mathrm{s}}(t)$ for each value of $v$.
We also find that the spin current vanishes as $v$ goes to zero.
That is what we expected since
there is no spin current generation when the shift vector is zero, 
as mentioned in Sec.~\ref{subsec:Orign_spin_dep}.

\section*{\label{sec:jc_E}S4: Charge current dependence on electric field}
\begin{figure}[h]
\subfigure[]{%
\includegraphics[bb=2 2 511 391,width=0.45\columnwidth]{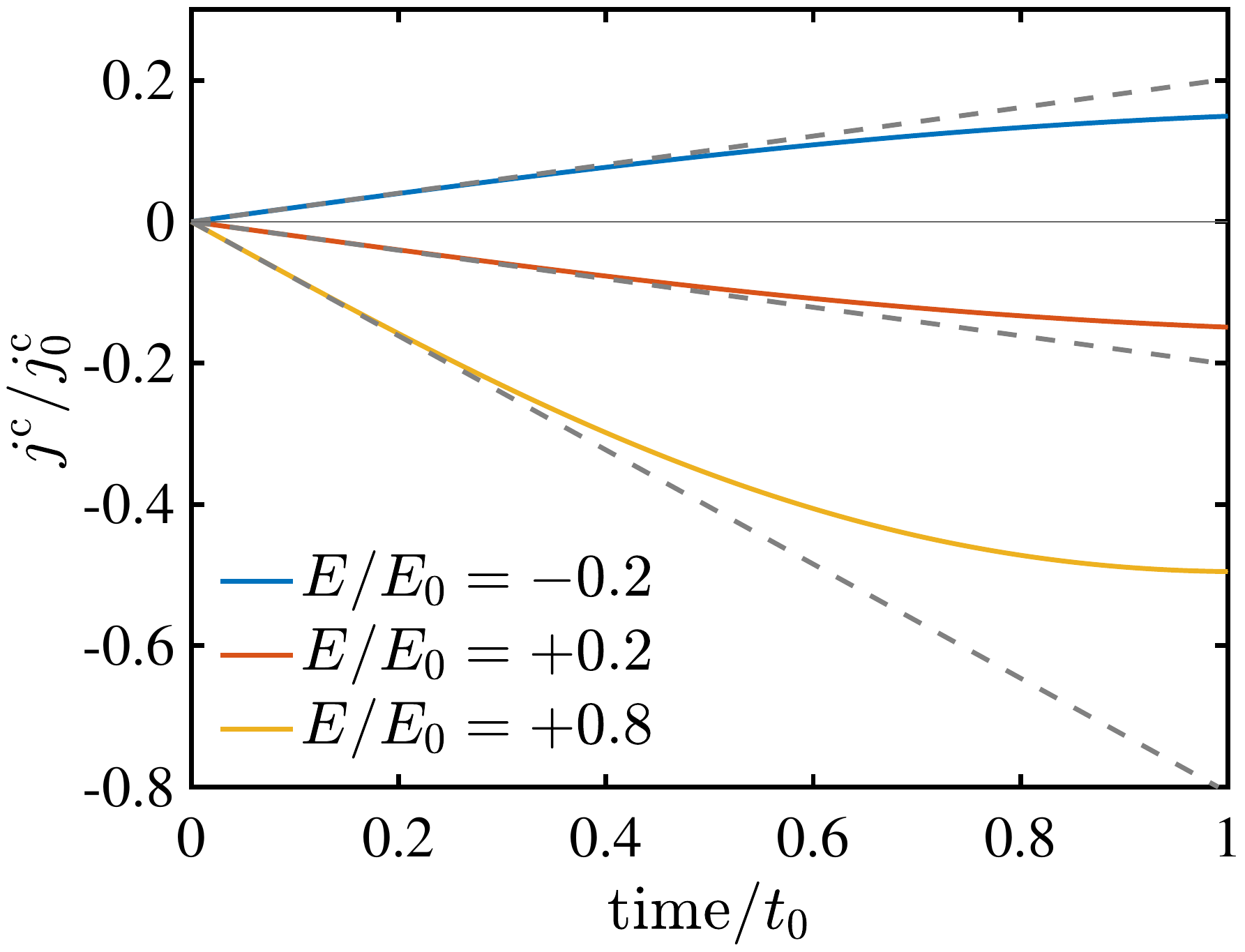}
\label{fig_jc_dE_PandT}
}%
\quad
\subfigure[]{%
\includegraphics[bb=2 2 511 391,width=0.45\columnwidth]{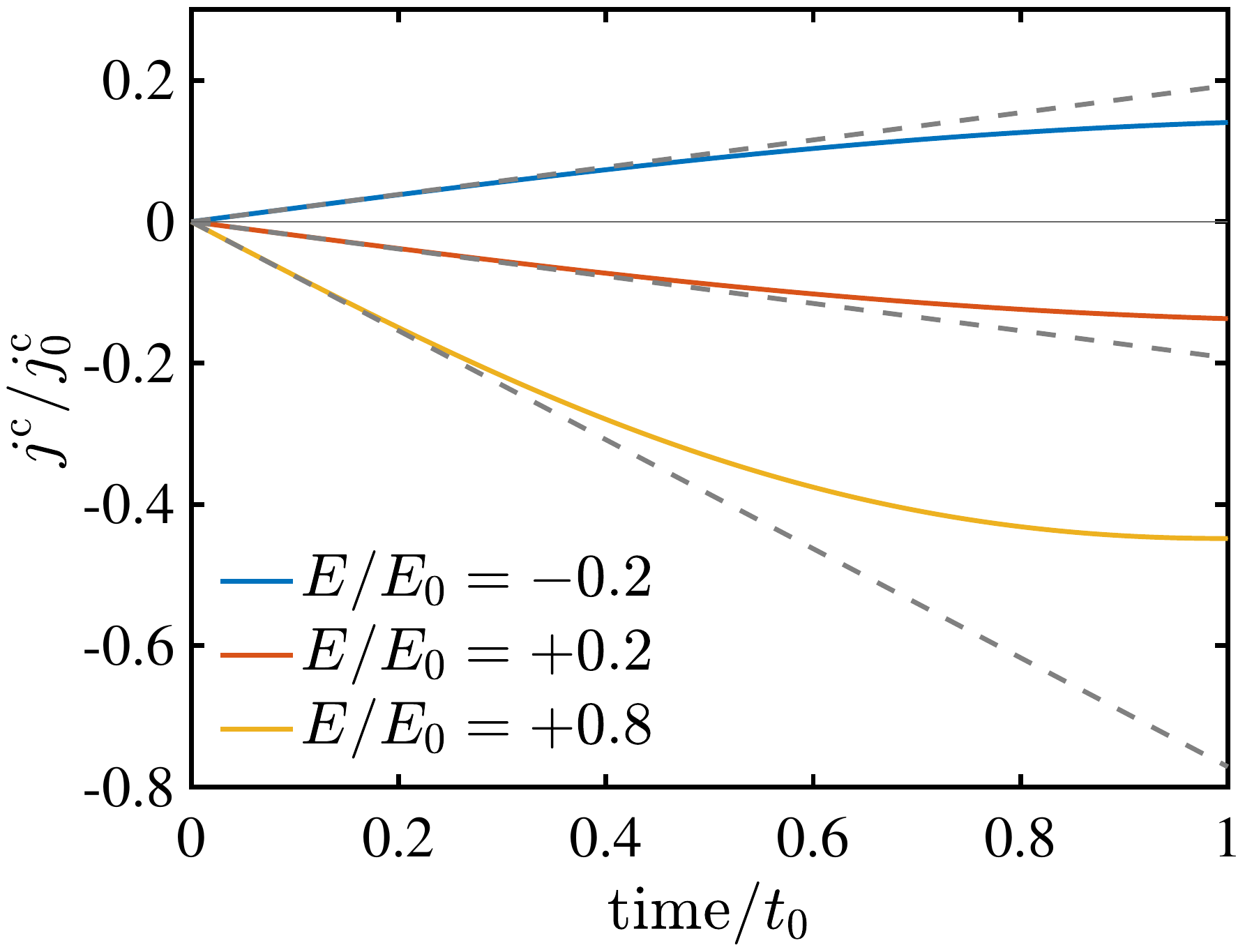}
\label{fig_jc_dE_PTsym}
}%
\caption{(Color online) 
Electric field dependence of the charge current $j^{\mathrm{c}}(E,t)$
shortly after the application of an electric field.
In both panels (a) and (b), 
the dashed curves represent $j^{\mathrm{c}}(E, t)$ from the analytical formula (\ref{eq_approx_jc}).
(a) $j^{\mathrm{c}}(E,t)$ for the model 
with inversion and time-reversal symmetries.
We used the same parameters $(u/{w}, v/{w})=(0.7, 0.4)$ as in Fig.~\ref{fig_js_dE}.
(b) $j^{\mathrm{c}}(E,t)$ for the PT symmetric model.
We used the same parameters $(u/{w}, v/{w}, \phi_s/{w})=(0.7, 0.4, 0.2)$ as in Fig.~\ref{fig_js_dE_PTsym}.
}
\label{fig_jc_dE}
\end{figure}

We touch on the behavior of the charge current in this section. 
Figures~\ref{fig_jc_dE} show the behaviors for
the two models in Sec.~\ref{zigzag_PandT} and Sec.~\ref{zigzag_PTsym}.
In each figure, we find that the analytical formula (\ref{eq_approx_jc})
fits the behavior of $j^{\mathrm{c}}(E, t)$ well.
This result supports the validity of the formula derived in
Appendix \ref{sec_approx_tunneling_current}.



\begin{thebibliography}{29}%
\makeatletter
\providecommand \@ifxundefined [1]{%
 \@ifx{#1\undefined}
}%
\providecommand \@ifnum [1]{%
 \ifnum #1\expandafter \@firstoftwo
 \else \expandafter \@secondoftwo
 \fi
}%
\providecommand \@ifx [1]{%
 \ifx #1\expandafter \@firstoftwo
 \else \expandafter \@secondoftwo
 \fi
}%
\providecommand \natexlab [1]{#1}%
\providecommand \enquote  [1]{``#1''}%
\providecommand \bibnamefont  [1]{#1}%
\providecommand \bibfnamefont [1]{#1}%
\providecommand \citenamefont [1]{#1}%
\providecommand \href@noop [0]{\@secondoftwo}%
\providecommand \href [0]{\begingroup \@sanitize@url \@href}%
\providecommand \@href[1]{\@@startlink{#1}\@@href}%
\providecommand \@@href[1]{\endgroup#1\@@endlink}%
\providecommand \@sanitize@url [0]{\catcode `\\12\catcode `\$12\catcode
  `\&12\catcode `\#12\catcode `\^12\catcode `\_12\catcode `\%12\relax}%
\providecommand \@@startlink[1]{}%
\providecommand \@@endlink[0]{}%
\providecommand \url  [0]{\begingroup\@sanitize@url \@url }%
\providecommand \@url [1]{\endgroup\@href {#1}{\urlprefix }}%
\providecommand \urlprefix  [0]{URL }%
\providecommand \Eprint [0]{\href }%
\providecommand \doibase [0]{https://doi.org/}%
\providecommand \selectlanguage [0]{\@gobble}%
\providecommand \bibinfo  [0]{\@secondoftwo}%
\providecommand \bibfield  [0]{\@secondoftwo}%
\providecommand \translation [1]{[#1]}%
\providecommand \BibitemOpen [0]{}%
\providecommand \bibitemStop [0]{}%
\providecommand \bibitemNoStop [0]{.\EOS\space}%
\providecommand \EOS [0]{\spacefactor3000\relax}%
\providecommand \BibitemShut  [1]{\csname bibitem#1\endcsname}%
\let\auto@bib@innerbib\@empty
\bibitem [{\citenamefont {Fabian}\ \emph {et~al.}(2007)\citenamefont {Fabian},
  \citenamefont {Matos-Abiague}, \citenamefont {Ertler}, \citenamefont
  {Stano},\ and\ \citenamefont {{\v{Z}}uti{\'{c}}}}]{Fabian2007}%
  \BibitemOpen
  \bibfield  {author} {\bibinfo {author} {\bibfnamefont {J.}~\bibnamefont
  {Fabian}}, \bibinfo {author} {\bibfnamefont {A.}~\bibnamefont
  {Matos-Abiague}}, \bibinfo {author} {\bibfnamefont {C.}~\bibnamefont
  {Ertler}}, \bibinfo {author} {\bibfnamefont {P.}~\bibnamefont {Stano}},\ and\
  \bibinfo {author} {\bibfnamefont {I.}~\bibnamefont {{\v{Z}}uti{\'{c}}}},\
  }\bibfield  {title} {\bibinfo {title} {{Semiconductor spintronics}},\
  }\href@noop {} {\bibfield  {journal} {\bibinfo  {journal} {Acta Physica
  Slovaca}\ }\textbf {\bibinfo {volume} {57}},\ \bibinfo {pages} {565}
  (\bibinfo {year} {2007})},\ \Eprint {https://arxiv.org/abs/0711.1461}
  {arXiv:0711.1461} \BibitemShut {NoStop}%
\bibitem [{\citenamefont {Culcer}\ and\ \citenamefont
  {Winkler}(2007)}]{Culcer2007a}%
  \BibitemOpen
  \bibfield  {author} {\bibinfo {author} {\bibfnamefont {D.}~\bibnamefont
  {Culcer}}\ and\ \bibinfo {author} {\bibfnamefont {R.}~\bibnamefont
  {Winkler}},\ }\bibfield  {title} {\bibinfo {title} {{Generation of spin
  currents and spin densities in systems with reduced symmetry}},\ }\href
  {https://doi.org/10.1103/PhysRevLett.99.226601} {\bibfield  {journal}
  {\bibinfo  {journal} {Physical Review Letters}\ }\textbf {\bibinfo {volume}
  {99}},\ {226601} (\bibinfo {year} {2007})}\BibitemShut {NoStop}%
\bibitem [{\citenamefont {Seemann}\ \emph {et~al.}(2015)\citenamefont
  {Seemann}, \citenamefont {K\"odderitzsch}, \citenamefont {Wimmer},\ and\
  \citenamefont {Ebert}}]{Seemann2015}%
  \BibitemOpen
  \bibfield  {author} {\bibinfo {author} {\bibfnamefont {M.}~\bibnamefont
  {Seemann}}, \bibinfo {author} {\bibfnamefont {D.}~\bibnamefont
  {K\"odderitzsch}}, \bibinfo {author} {\bibfnamefont {S.}~\bibnamefont
  {Wimmer}},\ and\ \bibinfo {author} {\bibfnamefont {H.}~\bibnamefont
  {Ebert}},\ }\bibfield  {title} {\bibinfo {title} {Symmetry-imposed shape of
  linear response tensors},\ }\href
  {https://doi.org/10.1103/PhysRevB.92.155138} {\bibfield  {journal} {\bibinfo
  {journal} {Physical Review B}\ }\textbf {\bibinfo {volume} {92}},\ \bibinfo
  {pages} {155138} (\bibinfo {year} {2015})}\BibitemShut {NoStop}%
\bibitem [{\citenamefont {Hamamoto}\ \emph {et~al.}(2017)\citenamefont
  {Hamamoto}, \citenamefont {Ezawa}, \citenamefont {Kim}, \citenamefont
  {Morimoto},\ and\ \citenamefont {Nagaosa}}]{Hamamoto2017}%
  \BibitemOpen
  \bibfield  {author} {\bibinfo {author} {\bibfnamefont {K.}~\bibnamefont
  {Hamamoto}}, \bibinfo {author} {\bibfnamefont {M.}~\bibnamefont {Ezawa}},
  \bibinfo {author} {\bibfnamefont {K.~W.}\ \bibnamefont {Kim}}, \bibinfo
  {author} {\bibfnamefont {T.}~\bibnamefont {Morimoto}},\ and\ \bibinfo
  {author} {\bibfnamefont {N.}~\bibnamefont {Nagaosa}},\ }\bibfield  {title}
  {\bibinfo {title} {{Nonlinear spin current generation in noncentrosymmetric
  spin-orbit coupled systems}},\ }\href
  {https://doi.org/10.1103/PhysRevB.95.224430} {\bibfield  {journal} {\bibinfo
  {journal} {Physical Review B}\ }\textbf {\bibinfo {volume} {95}},\ 
  {224430} (\bibinfo {year} {2017})}\BibitemShut {NoStop}%
\bibitem [{\citenamefont {Kitamura}\ \emph
  {et~al.}(2020{\natexlab{a}})\citenamefont {Kitamura}, \citenamefont
  {Nagaosa},\ and\ \citenamefont {Morimoto}}]{Kitamura2020b}%
  \BibitemOpen
  \bibfield  {author} {\bibinfo {author} {\bibfnamefont {S.}~\bibnamefont
  {Kitamura}}, \bibinfo {author} {\bibfnamefont {N.}~\bibnamefont {Nagaosa}},\
  and\ \bibinfo {author} {\bibfnamefont {T.}~\bibnamefont {Morimoto}},\
  }\bibfield  {title} {\bibinfo {title} {{Current response of nonequilibrium
  steady states in the Landau-Zener problem: Nonequilibrium Green's function
  approach}},\ }\href {https://doi.org/10.1103/PhysRevB.102.245141} {\bibfield
  {journal} {\bibinfo  {journal} {Physical Review B}\ }\textbf {\bibinfo
  {volume} {102}},\ \bibinfo {pages} {245141} (\bibinfo {year}
  {2020}{\natexlab{a}})}\BibitemShut {NoStop}%
\bibitem [{\citenamefont {Murakami}\ \emph {et~al.}(2003)\citenamefont
  {Murakami}, \citenamefont {Nagaosa},\ and\ \citenamefont
  {Zhang}}]{Murakami2003}%
  \BibitemOpen
  \bibfield  {author} {\bibinfo {author} {\bibfnamefont {S.}~\bibnamefont
  {Murakami}}, \bibinfo {author} {\bibfnamefont {N.}~\bibnamefont {Nagaosa}},\
  and\ \bibinfo {author} {\bibfnamefont {S.-c.}\ \bibnamefont {Zhang}},\
  }\bibfield  {title} {\bibinfo {title} {Dissipationless quantum spin current
  at room temperature},\ }\href {https://doi.org/10.1126/science.1087128}
  {\bibfield  {journal} {\bibinfo  {journal} {Science}\ }\textbf {\bibinfo
  {volume} {301}},\ \bibinfo {pages} {1348} (\bibinfo {year}
  {2003})}\BibitemShut {NoStop}%
\bibitem [{\citenamefont {Zener}(1934)}]{Zener1934}%
  \BibitemOpen
  \bibfield  {author} {\bibinfo {author} {\bibfnamefont {C.}~\bibnamefont
  {Zener}},\ }\bibfield  {title} {\bibinfo {title} {A theory of the electrical
  breakdown of solid dielectrics},\ }\href
  {https://doi.org/10.1098/rspa.1934.0116} {\bibfield  {journal} {\bibinfo
  {journal} {Proceedings of the Royal Society of London. Series A}\ }\textbf
  {\bibinfo {volume} {145}},\ \bibinfo {pages} {523} (\bibinfo {year}
  {1934})}\BibitemShut {NoStop}%
\bibitem [{\citenamefont {Landau}(1932)}]{Landau1932}%
  \BibitemOpen
  \bibfield  {author} {\bibinfo {author} {\bibfnamefont {L.~D.}\ \bibnamefont
  {Landau}},\ }\bibfield  {title} {\bibinfo {title} {Zur theorie der
  energie\"{u}bertragung. {II}},\ }\href@noop {} {\bibfield  {journal}
  {\bibinfo  {journal} {Physics of the Soviet Union}\ }\textbf {\bibinfo
  {volume} {2}},\ \bibinfo {pages} {46} (\bibinfo {year} {1932})}\BibitemShut
  {NoStop}%
\bibitem [{\citenamefont {Zener}(1932)}]{Zener1932}%
  \BibitemOpen
  \bibfield  {author} {\bibinfo {author} {\bibfnamefont {C.}~\bibnamefont
  {Zener}},\ }\bibfield  {title} {\bibinfo {title} {Non-adiabatic crossing of
  energy levels},\ }\href {https://doi.org/10.1098/rspa.1932.0165} {\bibfield
  {journal} {\bibinfo  {journal} {Proceedings of the Royal Society of London.
  Series A}\ }\textbf {\bibinfo {volume} {137}},\ \bibinfo {pages} {696}
  (\bibinfo {year} {1932})}\BibitemShut {NoStop}%
\bibitem [{\citenamefont {Landau}\ and\ \citenamefont
  {Lifshitz}(1981)}]{LandauLifshitzQM}%
  \BibitemOpen
  \bibfield  {author} {\bibinfo {author} {\bibfnamefont {L.~D.}\ \bibnamefont
  {Landau}}\ and\ \bibinfo {author} {\bibfnamefont {E.~M.}\ \bibnamefont
  {Lifshitz}},\ }\href@noop {} {\emph {\bibinfo {title} {Quantum Mechanics:
  Non-Relativistic Theory}}},\ Course of Theoretical Physics\ (\bibinfo
  {publisher} {Elsevier},\ \bibinfo {address} {Amsterdam},\ \bibinfo {year}
  {1981})\BibitemShut {NoStop}%
\bibitem [{\citenamefont {Zhang}\ \emph {et~al.}(2014)\citenamefont {Zhang},
  \citenamefont {Liu}, \citenamefont {Luo}, \citenamefont {Freeman},\ and\
  \citenamefont {Zunger}}]{Zhang2014a}%
  \BibitemOpen
  \bibfield  {author} {\bibinfo {author} {\bibfnamefont {X.}~\bibnamefont
  {Zhang}}, \bibinfo {author} {\bibfnamefont {Q.}~\bibnamefont {Liu}}, \bibinfo
  {author} {\bibfnamefont {J.-W.}\ \bibnamefont {Luo}}, \bibinfo {author}
  {\bibfnamefont {A.~J.}\ \bibnamefont {Freeman}},\ and\ \bibinfo {author}
  {\bibfnamefont {A.}~\bibnamefont {Zunger}},\ }\bibfield  {title} {\bibinfo
  {title} {{Hidden spin polarization in inversion-symmetric bulk crystals}},\
  }\href {https://doi.org/10.1038/nphys2933} {\bibfield  {journal} {\bibinfo
  {journal} {Nature Physics}\ }\textbf {\bibinfo {volume} {10}},\ \bibinfo
  {pages} {387} (\bibinfo {year} {2014})}\BibitemShut {NoStop}%
\bibitem [{\citenamefont {Yuan}\ \emph {et~al.}(2019)\citenamefont {Yuan},
  \citenamefont {Liu}, \citenamefont {Zhang}, \citenamefont {Luo},
  \citenamefont {Li},\ and\ \citenamefont {Zunger}}]{Yuan2019}%
  \BibitemOpen
  \bibfield  {author} {\bibinfo {author} {\bibfnamefont {L.}~\bibnamefont
  {Yuan}}, \bibinfo {author} {\bibfnamefont {Q.}~\bibnamefont {Liu}}, \bibinfo
  {author} {\bibfnamefont {X.}~\bibnamefont {Zhang}}, \bibinfo {author}
  {\bibfnamefont {J.-W.}\ \bibnamefont {Luo}}, \bibinfo {author} {\bibfnamefont
  {S.-S.}\ \bibnamefont {Li}},\ and\ \bibinfo {author} {\bibfnamefont
  {A.}~\bibnamefont {Zunger}},\ }\bibfield  {title} {\bibinfo {title}
  {{Uncovering and tailoring hidden Rashba spin-orbit splitting in
  centrosymmetric crystals}},\ }\href
  {https://doi.org/10.1038/s41467-019-08836-4}
  {\bibfield  {journal} {\bibinfo  {journal}
  {Nature Communications}\ }\textbf {\bibinfo {volume} {10}},\ {906}
  (\bibinfo {year} {2019})}\BibitemShut {NoStop}%
\bibitem [{\citenamefont {Berry}(1990)}]{Berry1990}%
  \BibitemOpen
  \bibfield  {author} {\bibinfo {author} {\bibfnamefont {M.~V.}\ \bibnamefont
  {Berry}},\ }\bibfield  {title} {\bibinfo {title} {{Geometric amplitude
  factors in adiabatic quantum transitions}},\ }\href
  {https://doi.org/10.1098/rspa.1990.0096} {\bibfield  {journal} {\bibinfo
  {journal} {Proceedings of the Royal Society of London. Series A: Mathematical
  and Physical Sciences}\ }\textbf {\bibinfo {volume} {430}},\ \bibinfo {pages}
  {405} (\bibinfo {year} {1990})}\BibitemShut {NoStop}%
\bibitem [{\citenamefont {Kitamura}\ \emph
  {et~al.}(2020{\natexlab{b}})\citenamefont {Kitamura}, \citenamefont
  {Nagaosa},\ and\ \citenamefont {Morimoto}}]{Kitamura2020a}%
  \BibitemOpen
  \bibfield  {author} {\bibinfo {author} {\bibfnamefont {S.}~\bibnamefont
  {Kitamura}}, \bibinfo {author} {\bibfnamefont {N.}~\bibnamefont {Nagaosa}},\
  and\ \bibinfo {author} {\bibfnamefont {T.}~\bibnamefont {Morimoto}},\
  }\bibfield  {title} {\bibinfo {title} {{Nonreciprocal Landau-Zener
  tunneling}},\ }\href
  {https://doi.org/10.1038/s42005-020-0328-0} {\bibfield  {journal} {\bibinfo  
  {journal} {Communications
  Physics}\ }\textbf {\bibinfo {volume} {3}},\ {63} (\bibinfo {year} 
  {2020}{\natexlab{b}})}\BibitemShut {NoStop}%
\bibitem [{\citenamefont {Takayoshi}\ \emph {et~al.}(2021)\citenamefont
  {Takayoshi}, \citenamefont {Wu},\ and\ \citenamefont {Oka}}]{Takayoshi2020}%
  \BibitemOpen
  \bibfield  {author} {\bibinfo {author} {\bibfnamefont {S.}~\bibnamefont
  {Takayoshi}}, \bibinfo {author} {\bibfnamefont {J.}~\bibnamefont {Wu}},\ and\
  \bibinfo {author} {\bibfnamefont {T.}~\bibnamefont {Oka}},\ }\href@noop {}
  {\bibinfo {title} {New aspects of nonadiabatic geometric effects --
  application to twisted Schwinger effect in Dirac and Weyl fermions}}
  (\bibinfo {year} {2021}),\ \Eprint {https://arxiv.org/abs/2005.01755}
  {arXiv:2005.01755} \BibitemShut {NoStop}%
\bibitem [{\citenamefont {Sipe}\ and\ \citenamefont
  {Shkrebtii}(2000)}]{Sipe2000}%
  \BibitemOpen
  \bibfield  {author} {\bibinfo {author} {\bibfnamefont {J.~E.}\ \bibnamefont
  {Sipe}}\ and\ \bibinfo {author} {\bibfnamefont {A.~I.}\ \bibnamefont
  {Shkrebtii}},\ }\bibfield  {title} {\bibinfo {title} {Second-order optical
  response in semiconductors},\ }\href
  {https://doi.org/10.1103/PhysRevB.61.5337} {\bibfield  {journal} {\bibinfo
  {journal} {Physical Review B}\ }\textbf {\bibinfo {volume} {61}},\ \bibinfo
  {pages} {5337} (\bibinfo {year} {2000})}\BibitemShut {NoStop}%
\bibitem [{\citenamefont {Young}\ and\ \citenamefont
  {Rappe}(2012)}]{Young2012}%
  \BibitemOpen
  \bibfield  {author} {\bibinfo {author} {\bibfnamefont {S.~M.}\ \bibnamefont
  {Young}}\ and\ \bibinfo {author} {\bibfnamefont {A.~M.}\ \bibnamefont
  {Rappe}},\ }\bibfield  {title} {\bibinfo {title} {First principles
  calculation of the shift current photovoltaic effect in ferroelectrics},\
  }\href {https://doi.org/10.1103/PhysRevLett.109.116601} {\bibfield  {journal}
  {\bibinfo  {journal} {Physical Review Letters}\ }\textbf {\bibinfo {volume}
  {109}},\ \bibinfo {pages} {116601} (\bibinfo {year} {2012})}\BibitemShut
  {NoStop}%
\bibitem [{\citenamefont {Cook}\ \emph {et~al.}(2017)\citenamefont {Cook},
  \citenamefont {Fregoso}, \citenamefont {{De Juan}}, \citenamefont {Coh},\
  and\ \citenamefont {Moore}}]{Cook2017}%
  \BibitemOpen
  \bibfield  {author} {\bibinfo {author} {\bibfnamefont {A.~M.}\ \bibnamefont
  {Cook}}, \bibinfo {author} {\bibfnamefont {B.~M.}\ \bibnamefont {Fregoso}},
  \bibinfo {author} {\bibfnamefont {F.}~\bibnamefont {{De Juan}}}, \bibinfo
  {author} {\bibfnamefont {S.}~\bibnamefont {Coh}},\ and\ \bibinfo {author}
  {\bibfnamefont {J.~E.}\ \bibnamefont {Moore}},\ }\bibfield  {title} {\bibinfo
  {title} {{Design principles for shift current photovoltaics}},\ }
   \href {https://doi.org/10.1038/ncomms14176}{\bibfield
  {journal} {\bibinfo  {journal} {Nature Communications}\ }\textbf {\bibinfo
  {volume} {8}},\ {14176} (\bibinfo {year} {2017})}\BibitemShut {NoStop}%
\bibitem [{\citenamefont {Morimoto}\ and\ \citenamefont
  {Nagaosa}(2016)}]{Morimoto2016}%
  \BibitemOpen
  \bibfield  {author} {\bibinfo {author} {\bibfnamefont {T.}~\bibnamefont
  {Morimoto}}\ and\ \bibinfo {author} {\bibfnamefont {N.}~\bibnamefont
  {Nagaosa}},\ }\bibfield  {title} {\bibinfo {title} {{Topological nature of
  nonlinear optical effects in solids}},\ }\href
  {https://doi.org/10.1126/sciadv.1501524} {\bibfield  {journal} {\bibinfo
  {journal} {Science Advances}\ }\textbf {\bibinfo {volume} {2}},\ {e1501524} 
  (\bibinfo {year} {2016})}\BibitemShut {NoStop}%
\bibitem [{\citenamefont {Kusunose}(2019)}]{Kusunose2019}%
  \BibitemOpen
  \bibfield  {author} {\bibinfo {author} {\bibfnamefont {H.}~\bibnamefont
  {Kusunose}},\ }\href@noop {} {\emph {\bibinfo {title} {Electron Theory of
  Spin-Orbit-Coupled Physics}}}\ (\bibinfo  {publisher} {Kodansha Ltd.},\
  \bibinfo {address} {Tokyo},\ \bibinfo {year} {2019})\BibitemShut {NoStop}%
\bibitem [{\citenamefont {Yanase}(2014)}]{Yanase2014}%
  \BibitemOpen
  \bibfield  {author} {\bibinfo {author} {\bibfnamefont {Y.}~\bibnamefont
  {Yanase}},\ }\bibfield  {title} {\bibinfo {title} {{Magneto-electric effect
  in three-dimensional coupled zigzag chains}},\ }\href
  {https://doi.org/10.7566/JPSJ.83.014703} {\bibfield  {journal} {\bibinfo
  {journal} {Journal of the Physical Society of Japan}\ }\textbf {\bibinfo
  {volume} {83}},\ {014703} (\bibinfo {year} {2014})}\BibitemShut
  {NoStop}%
\bibitem [{\citenamefont {Sugita}\ \emph {et~al.}(2015)\citenamefont {Sugita},
  \citenamefont {Hayami},\ and\ \citenamefont {Motome}}]{Sugita2015}%
  \BibitemOpen
  \bibfield  {author} {\bibinfo {author} {\bibfnamefont {Y.}~\bibnamefont
  {Sugita}}, \bibinfo {author} {\bibfnamefont {S.}~\bibnamefont {Hayami}},\
  and\ \bibinfo {author} {\bibfnamefont {Y.}~\bibnamefont {Motome}},\
  }\bibfield  {title} {\bibinfo {title} {{Antisymmetric spin-orbit coupling in
  a d-p model on a zigzag chain}},\ }\href
  {https://doi.org/10.1016/j.phpro.2015.12.051} {\bibfield  {journal} {\bibinfo
   {journal} {Physics Procedia}\ }\textbf {\bibinfo {volume} {75}},\ \bibinfo
  {pages} {419} (\bibinfo {year} {2015})}\BibitemShut {NoStop}%
\bibitem [{\citenamefont {Hayami}\ \emph {et~al.}(2016)\citenamefont {Hayami},
  \citenamefont {Kusunose},\ and\ \citenamefont {Motome}}]{Hayami2016}%
  \BibitemOpen
  \bibfield  {author} {\bibinfo {author} {\bibfnamefont {S.}~\bibnamefont
  {Hayami}}, \bibinfo {author} {\bibfnamefont {H.}~\bibnamefont {Kusunose}},\
  and\ \bibinfo {author} {\bibfnamefont {Y.}~\bibnamefont {Motome}},\
  }\bibfield  {title} {\bibinfo {title} {{Emergent spin-valley-orbital physics
  by spontaneous parity breaking}},\ }\href {https://doi.org/10.1088/0953-8984/28/39/395601}
  {\bibfield  {journal} {\bibinfo  {journal}
  {Journal of Physics Condensed Matter}\ }\textbf {\bibinfo {volume} {28}},\ {395601} 
  (\bibinfo {year} {2016})} \BibitemShut {NoStop}%
\bibitem [{\citenamefont {Hayami}\ \emph {et~al.}(2015)\citenamefont {Hayami},
  \citenamefont {Kusunose},\ and\ \citenamefont {Motome}}]{Hayami2015}%
  \BibitemOpen
  \bibfield  {author} {\bibinfo {author} {\bibfnamefont {S.}~\bibnamefont
  {Hayami}}, \bibinfo {author} {\bibfnamefont {H.}~\bibnamefont {Kusunose}},\
  and\ \bibinfo {author} {\bibfnamefont {Y.}~\bibnamefont {Motome}},\
  }\bibfield  {title} {\bibinfo {title} {{Spontaneous multipole ordering by
  local parity mixing}},\ }\href {https://doi.org/10.7566/JPSJ.84.064717}
  {\bibfield  {journal} {\bibinfo  {journal} {Journal of the Physical Society
  of Japan}\ }\textbf {\bibinfo {volume} {84}},\ {064717} (\bibinfo
  {year} {2015})}\BibitemShut {NoStop}%
\bibitem [{\citenamefont {Aoki}\ \emph {et~al.}(2014)\citenamefont {Aoki},
  \citenamefont {Tsuji}, \citenamefont {Eckstein}, \citenamefont {Kollar},
  \citenamefont {Oka},\ and\ \citenamefont {Werner}}]{Aoki2014}%
  \BibitemOpen
  \bibfield  {author} {\bibinfo {author} {\bibfnamefont {H.}~\bibnamefont
  {Aoki}}, \bibinfo {author} {\bibfnamefont {N.}~\bibnamefont {Tsuji}},
  \bibinfo {author} {\bibfnamefont {M.}~\bibnamefont {Eckstein}}, \bibinfo
  {author} {\bibfnamefont {M.}~\bibnamefont {Kollar}}, \bibinfo {author}
  {\bibfnamefont {T.}~\bibnamefont {Oka}},\ and\ \bibinfo {author}
  {\bibfnamefont {P.}~\bibnamefont {Werner}},\ } \bibfield  {title} {\bibinfo
  {title} {{Nonequilibrium dynamical mean-field theory and its applications}},\
  }\href{https://doi.org/10.1103/RevModPhys.86.779} {\bibfield  {journal} 
  {\bibinfo  {journal} {Reviews of Modern Physics}\
  }\textbf {\bibinfo {volume} {86}},\ {776} (\bibinfo {year} {2014})}
  \BibitemShut {NoStop}%
\bibitem [{\citenamefont {G{\"{o}}hler}\ \emph {et~al.}(2011)\citenamefont
  {G{\"{o}}hler}, \citenamefont {Hamelbeck}, \citenamefont {Markus},
  \citenamefont {Kettner}, \citenamefont {Hanne}, \citenamefont {Vager},
  \citenamefont {Naaman},\ and\ \citenamefont {Zacharias}}]{Gohler2011a}%
  \BibitemOpen
  \bibfield  {author} {\bibinfo {author} {\bibfnamefont {B.}~\bibnamefont
  {G{\"{o}}hler}}, \bibinfo {author} {\bibfnamefont {V.}~\bibnamefont
  {Hamelbeck}}, \bibinfo {author} {\bibfnamefont {T.~Z.}\ \bibnamefont
  {Markus}}, \bibinfo {author} {\bibfnamefont {M.}~\bibnamefont {Kettner}},
  \bibinfo {author} {\bibfnamefont {G.~F.}\ \bibnamefont {Hanne}}, \bibinfo
  {author} {\bibfnamefont {Z.}~\bibnamefont {Vager}}, \bibinfo {author}
  {\bibfnamefont {R.}~\bibnamefont {Naaman}},\ and\ \bibinfo {author}
  {\bibfnamefont {H.}~\bibnamefont {Zacharias}},\ }\bibfield  {title} {\bibinfo
  {title} {{Spin selectivity in electron transmission through self-assembled
  monolayers of double-stranded DNA}},\ }\href
  {https://doi.org/10.1126/science.1199339} {\bibfield  {journal} {\bibinfo
  {journal} {Science}\ }\textbf {\bibinfo {volume} {331}},\ \bibinfo {pages}
  {894} (\bibinfo {year} {2011})}\BibitemShut {NoStop}%
\bibitem [{\citenamefont {Xie}\ \emph {et~al.}(2011)\citenamefont {Xie},
  \citenamefont {Markus}, \citenamefont {Cohen}, \citenamefont {Vager},
  \citenamefont {Gutierrez},\ and\ \citenamefont {Naaman}}]{Xie2011}%
  \BibitemOpen
  \bibfield  {author} {\bibinfo {author} {\bibfnamefont {Z.}~\bibnamefont
  {Xie}}, \bibinfo {author} {\bibfnamefont {T.~Z.}\ \bibnamefont {Markus}},
  \bibinfo {author} {\bibfnamefont {S.~R.}\ \bibnamefont {Cohen}}, \bibinfo
  {author} {\bibfnamefont {Z.}~\bibnamefont {Vager}}, \bibinfo {author}
  {\bibfnamefont {R.}~\bibnamefont {Gutierrez}},\ and\ \bibinfo {author}
  {\bibfnamefont {R.}~\bibnamefont {Naaman}},\ }\bibfield  {title} {\bibinfo
  {title} {{Spin specific electron conduction through DNA oligomers}},\ }\href
  {https://doi.org/10.1021/nl2021637} {\bibfield  {journal} {\bibinfo
  {journal} {Nano Letters}\ }\textbf {\bibinfo {volume} {11}},\ \bibinfo
  {pages} {4652} (\bibinfo {year} {2011})}\BibitemShut {NoStop}%
\bibitem [{\citenamefont {Naaman}\ \emph {et~al.}(2019)\citenamefont {Naaman},
  \citenamefont {Paltiel},\ and\ \citenamefont {Waldeck}}]{Naaman2019a}%
  \BibitemOpen
  \bibfield  {author} {\bibinfo {author} {\bibfnamefont {R.}~\bibnamefont
  {Naaman}}, \bibinfo {author} {\bibfnamefont {Y.}~\bibnamefont {Paltiel}},\
  and\ \bibinfo {author} {\bibfnamefont {D.~H.}\ \bibnamefont {Waldeck}},\
  }\bibfield  {title} {\bibinfo {title} {{Chiral molecules and the electron
  spin}},\ }\href {https://doi.org/10.1038/s41570-019-0087-1} {\bibfield
  {journal} {\bibinfo  {journal} {Nature Reviews Chemistry}\ }\textbf {\bibinfo
  {volume} {3}},\ \bibinfo {pages} {250} (\bibinfo {year} {2019})}\BibitemShut
  {NoStop}%
\bibitem [{\citenamefont {Teschl}(2012)}]{Teschl2012}%
  \BibitemOpen
  \bibfield  {author} {\bibinfo {author} {\bibfnamefont {G.}~\bibnamefont
  {Teschl}},\ }\href@noop {} {\emph {\bibinfo {title} {Ordinary Differential
  Equations and Dynamical Systems}}},\ \bibinfo {series} {Graduate Studies in
  Mathematics}, Vol.\ \bibinfo {volume} {140}\ (\bibinfo  {publisher} {American
  Mathematical Soc.},\ \bibinfo {address} {Providence, Rhode Island},\ \bibinfo
  {year} {2012})\BibitemShut {NoStop}%
\end{thebibliography}

\end{document}